\documentclass[twoside,10pt]{article}
\usepackage{extsizes}
\usepackage[super,sort&compress,comma]{natbib} 
\usepackage[version=3]{mhchem}
\usepackage[b5paper, left=2.5cm, right=2.5cm, top=2.785cm, bottom=3.0cm]{geometry}
\usepackage{balance}
\usepackage{mathptmx}
\usepackage{sectsty}
\usepackage{graphicx} 
\usepackage{lastpage}
\usepackage[format=plain,justification=justified,singlelinecheck=false,font={stretch=1.125,small,sf},labelfont=bf,labelsep=space]{caption}
\usepackage{float}
\usepackage{fancyhdr}
\usepackage{fnpos}
\usepackage[english]{babel}
\addto{\captionsenglish}{%
  
}
\usepackage{array}
\usepackage{droidsans}
\usepackage{charter}
\usepackage[T1]{fontenc}
\usepackage[usenames,dvipsnames]{xcolor}
\usepackage{setspace}
\usepackage[compact]{titlesec}
\usepackage{hyperref}
\usepackage{amsmath,amssymb}
\usepackage{stackrel}

\usepackage{epstopdf}

\usepackage{bm}

\definecolor{cream}{RGB}{222,217,201}

\newcommand{\mytitle}{Ionic fluctuations in finite volumes: fractional noise and hyperuniformity}

\newcommand{\Lsim}{L_{\rm sim}}
\newcommand{\Lbox}{L_{\rm obs}}
\newcommand{\Lobs}{L_{\rm obs}}
\newcommand{\Robs}{R_{\rm obs}}

\newcommand{\dd}{\mathrm{d}}
\newcommand{\CN}{C_N}
\newcommand{\CQ}{C_Q}
\newcommand{\Vbox}{\mathcal{V}_{\rm obs}}
\newcommand{\kT}{k_BT}
\newcommand{\tdiff}{\tau_{\rm Diff}}
\newcommand{\tdeb}{\tau_{\rm Debye}}

\begin{document}

\pagestyle{fancy}
\thispagestyle{plain}
\fancypagestyle{plain}{
\renewcommand{\headrulewidth}{0pt}
}

\makeFNbottom
\makeatletter
\renewcommand\LARGE{\@setfontsize\LARGE{15pt}{17}}
\renewcommand\Large{\@setfontsize\Large{12pt}{14}}
\renewcommand\large{\@setfontsize\large{10pt}{12}}
\renewcommand\footnotesize{\@setfontsize\footnotesize{7pt}{10}}
\makeatother

\renewcommand{\thefootnote}{\fnsymbol{footnote}}
\renewcommand\footnoterule{\vspace*{1pt}%
\color{cream}\hrule width 3.5in height 0.4pt \color{black}\vspace*{5pt}} 
\setcounter{secnumdepth}{5}

\makeatletter 
\renewcommand\@biblabel[1]{#1}            
\renewcommand\@makefntext[1]%
{\noindent\makebox[0pt][r]{\@thefnmark\,}#1}
\makeatother 
\renewcommand{\figurename}{\small{Fig.}~}
\sectionfont{\sffamily\Large}
\subsectionfont{\normalsize}
\subsubsectionfont{\bf}
\setstretch{1.125} 
\setlength{\skip\footins}{0.8cm}
\setlength{\footnotesep}{0.25cm}
\setlength{\jot}{10pt}
\titlespacing*{\section}{0pt}{4pt}{4pt}
\titlespacing*{\subsection}{0pt}{15pt}{1pt}

\fancyfoot{}
\fancyfoot[RE,LO]{\vspace{-7.1pt}}
\fancyfoot[CO]{\vspace{-7.1pt}\hspace{13.2cm}}
\fancyfoot[CE]{\vspace{-7.1pt}\hspace{13.2cm}}
\fancyfoot[RO]{\footnotesize{\sffamily{1--\pageref{LastPage} ~\textbar  \hspace{2pt}\thepage}}}
\fancyfoot[LE]{\hspace{11.0cm} \footnotesize{\sffamily{1--\pageref{LastPage} ~\textbar  \hspace{2pt}\thepage}}}
\fancyhead{}
\renewcommand{\headrulewidth}{0pt} 
\renewcommand{\footrulewidth}{0pt}
\setlength{\arrayrulewidth}{1pt}
\setlength{\columnsep}{6.5mm}
\setlength\bibsep{1pt}

\makeatletter 
\newlength{\figrulesep} 
\setlength{\figrulesep}{0.5\textfloatsep} 

\newcommand{\topfigrule}{\vspace*{-1pt}%
\noindent{\color{cream}\rule[-\figrulesep]{\columnwidth}{1.5pt}} }

\newcommand{\botfigrule}{\vspace*{-2pt}%
\noindent{\color{cream}\rule[\figrulesep]{\columnwidth}{1.5pt}} }

\newcommand{\dblfigrule}{\vspace*{-1pt}%
\noindent{\color{cream}\rule[-\figrulesep]{\textwidth}{1.5pt}} }

\makeatother

  \begin{@twocolumnfalse}

\vspace{1em}
\sffamily
\noindent 

\noindent\LARGE{\textbf{\mytitle}} \\

 \noindent\large{Th\^{e} Hoang Ngoc Minh,\textit{$^{a}$} Benjamin Rotenberg,\textit{$^{a,b}$} and Sophie Marbach$^{\ast}$\textit{$^{a,c}$}} \\

\noindent\small{
Observing finite regions of a bigger system is a common experience, from microscopy to molecular simulations. In the latter especially, there is ongoing interest in predicting thermodynamic properties from tracking fluctuations in finite observation volumes. However, kinetic properties have received less attention, especially not in ionic solutions, where electrostatic interactions play a decisive role.
Here, we probe ionic fluctuations in finite volumes with Brownian dynamics and build an analytical framework that reproduces our simulation results and is broadly applicable to other systems with pairwise interactions.
Particle number and charge correlations exhibit a rich phenomenology with time, characterized by a diversity of timescales. The noise spectrum of both quantities decays as $1/f^{3/2}$, where $f$ is the frequency. This signature of fractional noise shows the universality of $1/f^{3/2}$ scalings when observing diffusing particles in finite domains. The hyperuniform behaviour of charge fluctuations, namely that correlations scale with the area of the observation volume, is preserved in time. Correlations even become proportional to the box perimeter at sufficiently long times. Our results pave the way to understand fluctuations in more complex systems, from nanopores to single-particle electrochemistry. 
} \\

 \end{@twocolumnfalse} 
 \vspace{0.6cm}


\renewcommand*\rmdefault{bch}\normalfont\upshape
\rmfamily
\section*{}
\vspace{-1cm}


\footnotetext{\textit{$^{a}$~Sorbonne Université, CNRS, Physicochimie des Électrolytes et Nanosystèmes Interfaciaux, F-75005 Paris,
France}}
\footnotetext{\textit{$^{b}$~Réseau sur le Stockage Electrochimique de l’Energie (RS2E), FR CNRS 3459, 80039 Amiens Cedex,
France }}
\footnotetext{\textit{$^{c}$~Courant Institute of Mathematical Sciences, New York University,
NY, 10012, U.S.A. }}
\footnotetext{$\ast$; E-mail: sophie@marbach.fr}

\begin{figure}[h]
\centering
  \includegraphics[width = \textwidth]{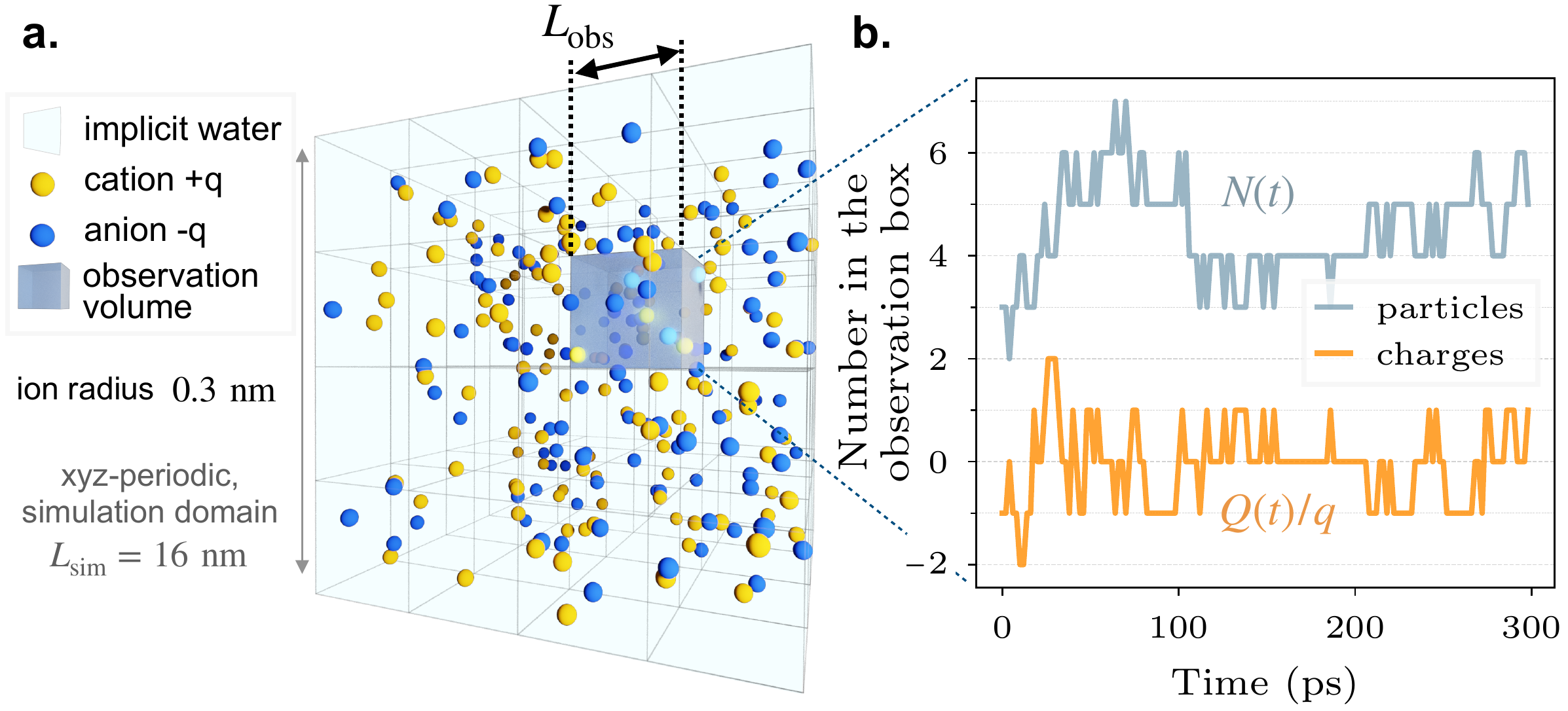}
  \caption{\textbf{Ionic fluctuations in finite observation volumes.} (a) Brownian dynamics simulation, here for a $C_0 = 52~\mathrm{mM}$ salt solution in a cubic, triply periodic simulation domain. Yellow (resp. blue) particles represent cations (resp. anions). The dark gray cube is a finite observation volume, here, one of the 64 boxes of size $\Lbox = \Lsim/4$. (b) Within the observation box, the number of particles $N = n_+ + n_-$ and  the charge $Q = q(n_+ - n_-)$ fluctuate, where $n_+$ (resp. $n_-$) is the number of positive (resp. negative) charges.}
  \label{fig:fig1}
\end{figure}

Microscopy techniques, from dynamic light scattering~\cite{berne2000dynamic} to fluorescence correlation spectroscopy~\cite{elson1974fluorescence}, generally rely on observing a small part of a much bigger underlying system. 
Understanding macroscopic properties from information at these scales, due in part to important fluctuations, is a major experimental challenge.
For this reason, there is a long history in theory and simulations, especially in molecular simulations, in tracking fluctuations on a small volume of an underlying larger simulation domain~\cite{schnell2011calculating,kruger2013kirkwood,chandler2005interfaces,kavokine2019ionic} (see also Fig.~\ref{fig:fig1}). The larger simulation domain serves as a reservoir for particles allowing one to probe behaviour within the grand canonical ensemble~\cite{kruger2013kirkwood,kavokine2019ionic,frenkel2001understanding}, without resorting to complex insertion/deletion rules~\cite{schnell2011calculating,frenkel2001understanding,robin2023ion,maginn2009molecular,belloni2019non}.

\textit{In the static limit}, fluctuations in finite observation volumes give access to various thermodynamic properties. Charge fluctuations in Coulombic systems can quantify screening properties~\cite{van1979thermodynamics,martin1980charge,lebowitz1983charge,levesque2000charge,bekiranov1998fluctuations,kim2005screening,kim2008charge,jancovici2003charge,kavokine2019ionic}. Fluctuations of water molecules in observation volumes near interfaces can probe surface hydrophobicity and solvation free energies~\cite{chandler2005interfaces,patel2010fluctuations,rotenberg2011molecular}. More generally, density fluctuations integrated over increasing volumes correspond, in the limit of infinite volumes, to Kirkwood-Buff integrals~\cite{kirkwood1951statistical} from which it is possible to extract various thermodynamic properties of the fluid such as partial molar volumes, compressibility, and chemical potentials~\cite{kirkwood1951statistical,kusalik1987thermodynamic,schnell2011calculating,kruger2013kirkwood,dawass2019kirkwood,cheng_computing_2022}.  

In contrast, resolving fluctuating \textit{dynamics} in finite observation volumes has received less attention. 
Yet, Green-Kubo integrals -- the kinetic counterpart of Kirkwood-Buff integrals -- give access to various dynamic properties. Integration of fluctuating fluxes over an entire domain enables to probe conductivity, permeance, friction on surfaces, and more,~\cite{bocquet2010nanofluidics,bocquet1994hydrodynamic,van2014statistical,guerin2015kubo,oga2019green,espanol2019solution,marbach2019osmosis,minh2022frequency,pireddu2022frequency,zorkot2018current,zorkot2016power,zorkot2016current,detcheverry2013thermal,sega_calculation_2013,cox_finite_2019,caillol_dielectric_1987,caillol_theoretical_1986}, including far from equilibrium~\cite{dal2019linear,lesnicki2020field,lesnicki2021molecular,chun2021nonequilibrium}. 
The relevance of Green-Kubo integrals over sub-volumes has only recently been raised, particularly to coarse-grain molecular dynamics near interfaces~\cite{duque2019discrete}. Numerous coarse-graining techniques, from Mori-Zwanzig approaches to (fluctuating) Lattice-Boltzmann or Fluctuating Hydrodynamics, which are crucial to access mesoscale dynamics~\cite{gubbiotti2022electroosmosis}, rely on a detailed understanding of fluctuations in finite volumes~\cite{dunweg_statistical_2007, dunweg_progress_2009, asta2017transient, parsa2017lattice, parsa2020large, tischler2022thermalized, schilling2022coarse,espanol2019solution,donev2010accuracy,donev_fluctuating_hydro_2019,peraud_fluctuating_2017}.

The study of \textit{ionic} fluctuations in finite volumes is especially intriguing. Without electrostatic interactions, particle number fluctuations at a steady state should scale like the average number of particles in the observation volume, $\sim \Lobs^3$ if $\Lbox$ is the corresponding size. However, fluctuations of the charge $Q$ are dramatically screened by electrostatics and scale with the \textit{area} of the observation volume for sufficiently large volumes, $\langle Q^2 \rangle \sim \Lbox^2$ ~\cite{martin1980charge,lebowitz1983charge,levesque2000charge,bekiranov1998fluctuations,kim2005screening,kim2008charge,jancovici2003charge} -- a generic phenomenon termed hyperuniformity~\cite{torquato2003local,torquato2018hyperuniform,ghosh2017fluctuations,leble2021two}. How does this screening behaviour pertain in time? 

Furthermore, dynamic features of electrolytes, such as conductivity, are not characterized by a universal timescale. 
An ion's self-diffusion coefficient $D$ -- related to its mobility -- and the Debye screening length $\lambda_D$ -- which quantifies the length scale of electrostatic interactions -- define the characteristic timescale $\tdeb = \lambda_D^2/D$ for the relaxation time of fluctuations in a bulk electrolyte. However, other length scales, such as, here, the size of the observation volume, $\Lobs$, provide other timescales, such as $\tdiff = \Lobs^2/D$, the time to diffuse across the finite volume. \textit{De facto}, various mixed timescales, as a combination $\lambda_D^{\nu} \Lobs^{2-\nu}/D$ with $\nu$ a real number, will also play a role, as was seen in the charging dynamics of two parallel plates separated by a distance $L$ playing the role of $\Lobs$ here~\cite{bazant2004diffuse,palaia2023charging,minh2022frequency}.
Which timescale dominates fluctuations in finite volumes?

Here, we use a combination of Brownian dynamics simulations and analytical calculations to rationalize ionic fluctuations in finite volumes (Sec.~\ref{sec:secmethodo}). We probe (see Fig.~\ref{fig:fig1}) the particle number $N$ and charge $Q$ in cubic observation volumes with side $\Lbox$, smaller than the overall system size, for an electrolyte with various concentrations, at equilibrium. The correlations in the particle number fluctuations $N$ decay algebraically in time and are not affected by electrostatics (Sec.~\ref{sec:2}). In contrast, charge fluctuations $Q$ strongly depend on electrostatic interactions (Sec.~\ref{sec:3}). In the static limit, we recover hyperuniformity when the observation volume is much larger than the Debye length $\Lobs \gg \lambda_D$. The dynamic response of charge correlations encompasses a rich phenomenology that depends on the separation of length scales. To lift this ambiguity, we introduce a global timescale, defined as a weighted integral over the structure factor (Eq.~\eqref{eq:T}), which quantifies the impact of either length scale on the relaxation time. The noise spectrum of both $Q$ and $N$ features a characteristic decay as $1/f^{3/2}$ where $f$ is the frequency, a signature of fractional noise~\cite{marbach2021intrinsic}, showing that such noise is a universal property of diffusing particles observed in finite volumes. The present framework can describe fluctuations in finite volumes for particles with different pairwise interactions, which allows us to discuss our results in the broader context of coarse-graining techniques, hyperuniformity, and electrochemical noise in confined geometries (Sec.~\ref{sec:4}).

\section{Methodological overview}
\label{sec:secmethodo}

\subsection{Numerical methods}

We perform Brownian dynamics (BD) simulations of model electrolyte solutions (see Fig.~\ref{fig:fig1}-a). Specifically, we solve overdamped Langevin equations to describe the stochastic ion motion in an implicit solvent
\begin{equation}
    \frac{d\bm{x}_i}{dt} = -\frac{D_i}{k_B T} \sum_{j\neq i} \bm{\nabla} V^{\rm Coul}_{ij} (||\bm{x}_i - \bm{x}_j||) + \sqrt{2D_i} \bm{\eta}_i(t)
    \label{eq:Langevin}
\end{equation}
where $\bm{x}_i$ is the 3D position of particle $i$, $D_i$ its diffusion coefficient, $k_B T$ the thermal energy and $\bm{\eta}_i$ a 3D gaussian white noise representing the action of the implicit solvent on the ions (such that $\langle \eta_{x,i}\rangle = 0$ and $\langle \eta_{x,i}(t)\eta_{x',i}(t') \rangle =  \delta(t-t')\delta_{xx'} \delta_{ij}$ where $x$ here indicates the $x^{\text{th}}$ component of the vector $\bm{\eta}_i$). 
$V_{ij}^{\mathrm{Coul}}$ corresponds to Coulomb interactions
\begin{equation}
    V_{ij}^{\mathrm{Coul}}(r=||\bm{x}_i - \bm{x}_j||) = \frac{q_i q_j}{4 \pi \epsilon_0 \epsilon_w r}
    \label{eq:COULOMB_POT}
\end{equation}
where $q_i$ is the charge of particle $i$, $\epsilon_0$ is the vacuum permittivity, and $\epsilon_w$ the relative permittivity of the fluid. To avoid ionic collapse, we also add pairwise short-range repulsive interactions (not specified in Eq.~\eqref{eq:Langevin}, see details in Appendix A). We take parameters to describe a typical symmetric salt solution, here KCl in water, broadly used in experiments~\cite{bocquet2010nanofluidics,marbach2019osmosis,secchi2016scaling,powell2009nonequilibrium,knowles2019noise,knowles2021current,smeets2008noise}: $q_{+} = e = -q_{-} \equiv q$ where $e$ is the elementary charge, $D_+ = D_- = D = 1.5\times 10^{-9}~\mathrm{m^2/s}$ and $\epsilon_\mathrm{w} = 78.5$. 
We conduct simulations with $N_{0} = N_+ = N_-$ ion pairs enclosed in a cubic simulation domain of side $\Lsim = 16~\mathrm{nm}$ and periodic boundary conditions. The salt concentration is, therefore, $C_0 = N_0/\Lsim^3$. Additional simulation details may be found in Appendix A.  

Here, we have chosen minimal interactions between ions; in particular, we have neglected hydrodynamic interactions~\cite{te_vrugt_classical_2020} to isolate the effect of electrostatic interactions. In a companion paper~\cite{FDspectra}, we found that the same BD simulations, in a $\sim1$~M aqueous electrolyte solution, compared with molecular dynamics simulations, capture well the main features of the dynamic structure factor of charges. 
However, deviations are observed for intermediate wavenumbers, which can be partially improved by improving the description of static correlations. Electrostatic and hydrodynamic interactions can be jointly addressed with either lengthy molecular dynamics or faster Brownian dynamics of ions with a fluctuating implicit solvent~\cite{ladiges1,ladiges2}. The study of hydrodynamic interactions (albeit in the absence of electrostatics) will be the focus of a further study~\cite{AliceExclusion} using fluctuating hydrodynamics~\cite{sprinkle2017large,hashemi2022computing}. 

Our goal in this work is to understand the defining features of the fluctuating number of particles $N=n_+(\Lbox)+n_-(\Lbox)$ and the charge $Q=q[n_+(\Lbox)-n_-(\Lbox)]$ in cubic observation volumes of side $\Lbox$ within the simulation domain~(see Fig.~\ref{fig:fig1}-a); where $n_{\pm}(\Lbox)$ refer to the number of positively (resp. negatively) charged particles in the observation volume. Recording particle positions, we find that the two quantities fluctuate in time (see Fig.~\ref{fig:fig1}-b) taking discrete values, either around the average box occupation $\langle N \rangle = 2C_0\Lbox^3$ or the average zero charge $\langle Q \rangle = 0$. To analyze their statistical properties, we examine their static $\CN(0), \CQ(0)$ and dynamic correlations, $\CN(t) = \langle N(t) N(0)\rangle - \langle N \rangle^2$ and $\CQ(t) = \langle Q(t) Q(0) \rangle$. Here, we have checked that both $N$ and $Q$ follow Gaussian distributions around their average values, which means relevant insight can be obtained without considering higher-order correlations. Yet, in different geometries, such as near interfaces, we might expect deviations from Gaussian distributions, which can be exploited to calculate other thermodynamic quantities (see \emph{e.g.} Refs.~\citenum{chandler2005interfaces, patel2010fluctuations} for the link between water density fluctuations and the hydrophobic/hydrophilic character of surfaces). 

\subsection{Stochastic Density Functional Theory}

To quantify particle statistics in boxes, we also rely on analytical calculations that introduce at the mean-field level the same physical ingredients as in the BD simulations. Let $C(\bm{x},t)$ and $\rho(\bm{x},t)$ be the number and charge density fields, respectively, so that the instantaneous number and charge in the observation volume $\Vbox$ are
\begin{equation}
    N(t) = \int_{\Vbox} C(\bm{x},t) \dd \bm{x}  \,\, \text{and}\,\, Q(t) = \int_{\Vbox} \rho(\bm{x},t) \dd \bm{x}
    \, .
\end{equation}
In the following, we also consider $c(\bm{x},t)=C(\bm{x},t)-2C_0$, the excess density relative to the mean background ion density $2C_0$. While there are diverse strategies to calculate the statistical properties of $N$ and $Q$~\cite{schnell2011calculating,marbach2021intrinsic,te_vrugt_classical_2020}, stochastic Density Functional Theory (sDFT)~\cite{dean1996langevin,kawasaki1994stochastic} stands out here for its simplicity. sDFT directly describes the fluctuations on the continuous fields $C(\bm{x},t)$ and $\rho(\bm{x},t)$ due to individual particle diffusion and has been successfully applied to electrolytes to recover Onsager relations~\cite{demery2016conductivity}. It is also especially suited to extract kinetic properties~\cite{jardat2022diffusion,mahdisoltani2021transient}. 
Starting from Poisson-Nernst-Planck equations, using sDFT to introduce fluctuations on individual particle fields, and then assuming fluctuations are small compared to the background density ($|c| \sim |\rho/q| \ll C_0$), the fields satisfy~\cite{mahdisoltani2021transient} 
\begin{equation}
\begin{cases}
    \partial_t c &= D \nabla^2 c + \sqrt{4 D C_0}\bm{\nabla} \cdot \bm{\eta}_c \\
    \partial_t \rho &= D \nabla^2 \rho - D \frac{1}{\lambda_D^2} \rho + \sqrt{4 D C_0}\bm{\nabla}\cdot  \bm{\eta}_{\rho} 
\end{cases}    
\label{eq:L-sPNP}
\end{equation}
where $\lambda_D = \sqrt{ \frac{\kT \epsilon_0 \epsilon_w }{q^2 (2C_0)} }$ is the Debye screening length, and the $\bm{\eta}_X$ (for $X\in\{c,\rho\}$) are 3D Gaussian white noises with uncorrelated components (see Appendix B for a short derivation).

Within this framework, we obtain the structure factors for the density ($X = c$) and the charge ($X = \rho$). They are best expressed in Fourier space. If $\tilde{X}(\bm{k},\omega)$ is the Fourier transform of $X(\bm{X},t)$\footnote{With the convention that $\tilde{X}(\bm{k},\omega) = \iint e^{-i\omega t} e^{-i \bm{k}\cdot \bm{x}} X(\bm{x},t) \dd \bm{x} \dd t$} then 
\begin{equation}
    \langle \tilde{X}(\bm{k},\omega) \tilde{X}(\bm{k}',\omega') \rangle =  2C_0 q_X^2 (2\pi)^4 S_{XX}(\bm{k},\omega) \delta(\omega+\omega') \delta^3(\bm{k}+ \bm{k}')
    \label{eq:S_fourier}
\end{equation}
where $q_{\rho}^2 = q^2 = (Ze)^2$, $q_{c}^2 = 1$, and $S_{XX}(\bm{k},\omega) $ is the structure factor. For both fields, we find
\begin{equation}
\begin{split}
    &S_{XX}(\bm{k},\omega) = \frac{2 D k^2}{\omega^2 + (Dk^2/S^{\rm static}_{XX}(k))^2} \\
    &\,\,\, \text{with static structure factors}\,\,\,
    S^{\rm static}_{cc}(k) = 1, \,\,\,\,  S^{\rm static}_{\rho\rho}(k) = \frac{k^2}{k^2 + \kappa_D^2} 
    \end{split}
    \label{eq:Sgeneral}
\end{equation}
where $k = |\bm{k}|$ and $\kappa_D = 1/\lambda_D$. Eq.~\ref{eq:Sgeneral} for the static charge structure factor $S_{\rho\rho}^{\mathrm{static}}(k)$ corresponds to the classical result of Debye-H\"uckel~\cite{hansen2013theory, FDspectra}, which is expected since linearized sDFT yields the lowest order coupling between diffusion and electrostatics, \textit{i.e.} dynamics close to equilibrium. More generally, the dynamic structure factor expression in Eq.~\eqref{eq:Sgeneral} holds for various fields, as long as they derive from Markovian (no memory) and gaussian processes (forces are conservative and derive from an energy that is quadratic in the field)~\cite{FDspectra,AliceExclusion,marbach2018transport}. The present formalism can thus easily be extended to study different pairwise interactions, such as steric interactions. Still, it should be modified to account \textit{e.g.} for hydrodynamic interactions~\cite{te_vrugt_classical_2020,AliceExclusion,zorkot2016current,sprinkle2017large,hashemi2022computing}. 

The correlations of $N$ and $Q$ are then simply given by 
\begin{equation}
    \begin{cases}
        &\CN(t) = \displaystyle \iint_{\Vbox} \langle c(\bm{x},t) c(\bm{x}',0) \rangle \dd \bm{x} \dd \bm{x}' = \Lbox^3 \iint \frac{\dd \bm{k}\dd \omega}{(2\pi)^4} e^{i\omega t} f_{\mathcal{V}}(\bm{k}) S_{cc}(k,\omega), \\
        &\CQ(t) = \displaystyle \iint_{\Vbox} \langle \rho(\bm{x},t) \rho(\bm{x}',0) \rangle \dd \bm{x} \dd \bm{x}' =\Lbox^3 \iint \frac{\dd \bm{k}\dd \omega}{(2\pi)^4} e^{i\omega t} f_{\mathcal{V}}(\bm{k}) S_{\rho\rho}(k,\omega), 
    \end{cases}
    \label{eq:deltaNS}
\end{equation}
where 
\begin{equation}
 f_\mathcal{V}(\bm{k}) =  \frac{1}{\Lbox^3} \iint \dd \bm{x} \dd \bm{x}' e^{i \bm{k}\cdot (\bm{x}- \bm{x}')}
 \label{eq:fv}
\end{equation}
is a geometrical volume factor that accounts for the shape of the observation box in Fourier space. Eq.~\eqref{eq:deltaNS} extends Kirkwood-Buff type integrals beyond the static regime~\cite{van1979thermodynamics,jancovici2003charge,kim2008charge,schnell2011calculating,kruger2013kirkwood,kusalik1987thermodynamic}. In addition, compared to previous calculations for the static case which were conducted in real space~\cite{van1979thermodynamics, jancovici2003charge, kim2008charge, schnell2011calculating, kruger2013kirkwood, kusalik1987thermodynamic}, calculations in Fourier space are more straightforward. 
Here, we will focus on cubic observation boxes where statistical analysis can be sped up. However, we expect our results to persist for other geometries, especially the scaling laws we unravel. We demonstrate the generality of these scalings in Appendix D, where we show analytically that they also apply to spherical observation volumes.

\section{Particle number correlations decay algebraically with time}
\label{sec:2}

\begin{figure}[h]
\centering
  \includegraphics[width = \textwidth]{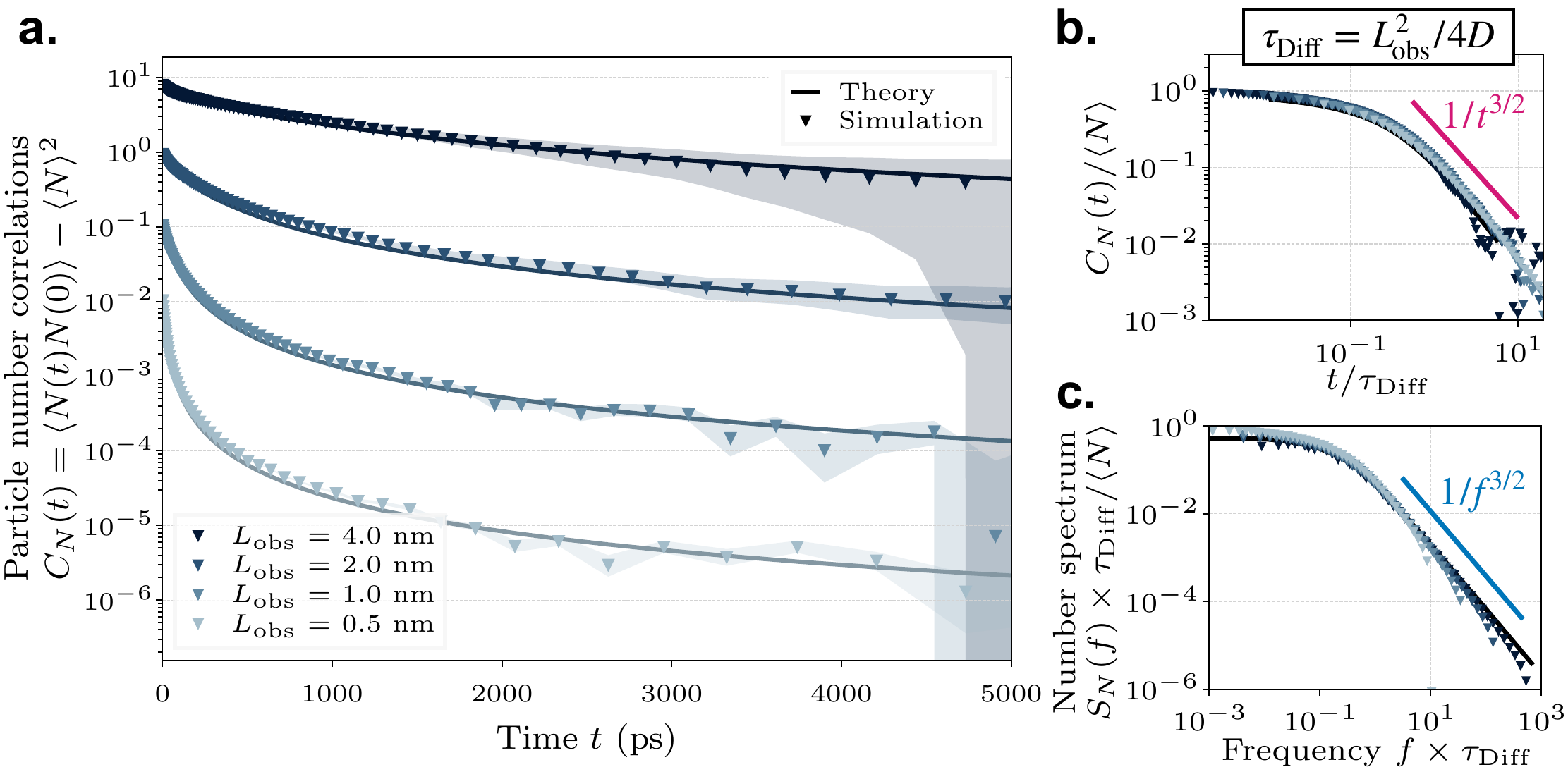}
  \caption{\textbf{Algebraic decay of particle number fluctuations.} (a) Particle number correlations with time, ranging from light blue to dark blue for increasing observation box size. Brownian dynamics results are shown as triangles, with shaded areas indicating one standard deviation around the mean, while lines are predictions from Eq.~\eqref{eq:deltaN}. (b) Rescaled (a) plot showing algebraic decay at long times as $1/t^{3/2}$. (c) Associated frequency spectrum with the $1/f^{3/2}$ signature of fractional noise~\cite{marbach2021intrinsic}. Here, $C_0 = 104~\mathrm{mM}$; colored legends are shared across (a-c).}
  \label{fig:fig2}
\end{figure}

We start by analyzing particle number correlations, $\CN(t) = \langle N(t) N(0)\rangle - \langle N \rangle^2$, and present BD simulations results for various observation box sizes in Fig.~\ref{fig:fig2}-a (triangles). We find that correlations decay with time due to particle exchanges between the observation box and the rest of the simulation domain. With larger observation boxes, correlations increase in magnitude since more particles participate in the fluctuations. To rationalize this behaviour, we use the sDFT framework. 
Inserting the expression of the spectrum Eq.~\eqref{eq:Sgeneral} in Eq.~\eqref{eq:deltaNS} we find after integration (see Appendix C), 
\begin{equation}
\begin{split}
    \CN(t) &= \langle N \rangle \left[ f_N\left(\frac{4Dt}{\Lbox^2} \right)  \right]^3,  \\
   & \text{where}\,\, f_N\left(\frac{t}{\tdiff} \equiv \frac{4D t}{\Lbox^2} \right) =   \sqrt{\frac{t}{\tdiff \pi}} \left( e^{-\tdiff/t} - 1\right) + \mathrm{erf} \left( \sqrt{\frac{\tdiff}{t}} \right). 
    \label{eq:deltaN}
    \end{split}
\end{equation}
We compare BD results with Eq.~\eqref{eq:deltaN} in Fig.~\ref{fig:fig2}-a (symbols and lines, respectively). The excellent agreement shows that sDFT is indeed well suited to predict particle number fluctuations. 

Eq.~\eqref{eq:deltaN} shows that number fluctuations scale with the average number of particles in the observation box, $\langle N \rangle$, and are determined by a single timescale $\tdiff = \Lbox^2/4D$ corresponding to particle diffusion across the observation box. This is confirmed in Fig.~\ref{fig:fig2}-b, which shows that all BD results collapse on a master curve, well described by Eq.~\eqref{eq:deltaN}, when rescaled by $\langle N \rangle$ and time by $\tau_{\rm Diff}$. Furthermore, we find that the correlations decay algebraically as $t^{-3/2}$ at long times. Expanding Eq.~\eqref{eq:deltaN} we find $\CN(t)/\langle N \rangle = (\tdiff/\pi t)^{3/2}$, which confirms the exponent of the algebraic decay. This \textit{slow} relaxation of the correlations indicates that particle rearrangements are slow with time due to their diffusive or Brownian nature.

Finally, the noise spectrum $S_N(f)$ associated with $N(t)$, reported in Fig.~\ref{fig:fig2}-c, decays at high frequencies as $1/f^{3/2}$, a signature of fractional noise. This $3/2$ exponent is not related to the long-time algebraic decay of the correlations; rather, it corresponds to the early time behaviour as $\CN(t)/\langle N \rangle \simeq 1 - \sqrt{t/\pi\tau_{\rm Diff}}$. 
Overall, the statistical properties of $N(t)$ are thus characteristic of a so-called fractional Brownian walk, with ``diffusion coefficient'' $\sqrt{1/\pi\tau_{\rm Diff}}$ and Hurst index H = 1/4~\cite{mandelbrot1968fractional,burov2011single}. 
The physical origin of this peculiar mathematical property comes from boundary crossings, here, that of the observation volume. Similar fractional noise signatures were predicted in 1D for Brownian particles with no interactions~\cite{marbach2021intrinsic}. Remarkably, this fractional feature pertains here in 3D, with particle interactions, showing that fractional noise is a universal property of Brownian motion, which arises as soon as a quantity involves particles crossing boundaries.   

Surprisingly, our results for particle number fluctuations do not depend on electrostatic properties. While this is somewhat expected at low enough salt concentrations $C_0$, steric effects should modify number fluctuations at high concentrations. Steric effects result in oscillations of the static structure factor~\cite{hansen2013theory,thorneywork2018structure}, that are only weakly captured by BD~\cite{FDspectra} and not at this stage with sDFT in Eq.~\eqref{eq:Sgeneral}. Steric effects can, however, be captured by improving the expression of the static structure factor analytically~\cite{thorneywork2018structure,hansen1982rescaled,dufreche2002ionic} or by fitting numerically obtained structure factors~\cite{FDspectra} and will be the object of future work~\cite{AliceExclusion}.

\section{Exotic signatures in charge fluctuations}
\label{sec:3}

We now turn to charge fluctuations within finite observation volumes. 

\subsection{Hyperuniformity in the static regime}

\begin{figure}[h]
\centering
  \includegraphics[width = \textwidth]{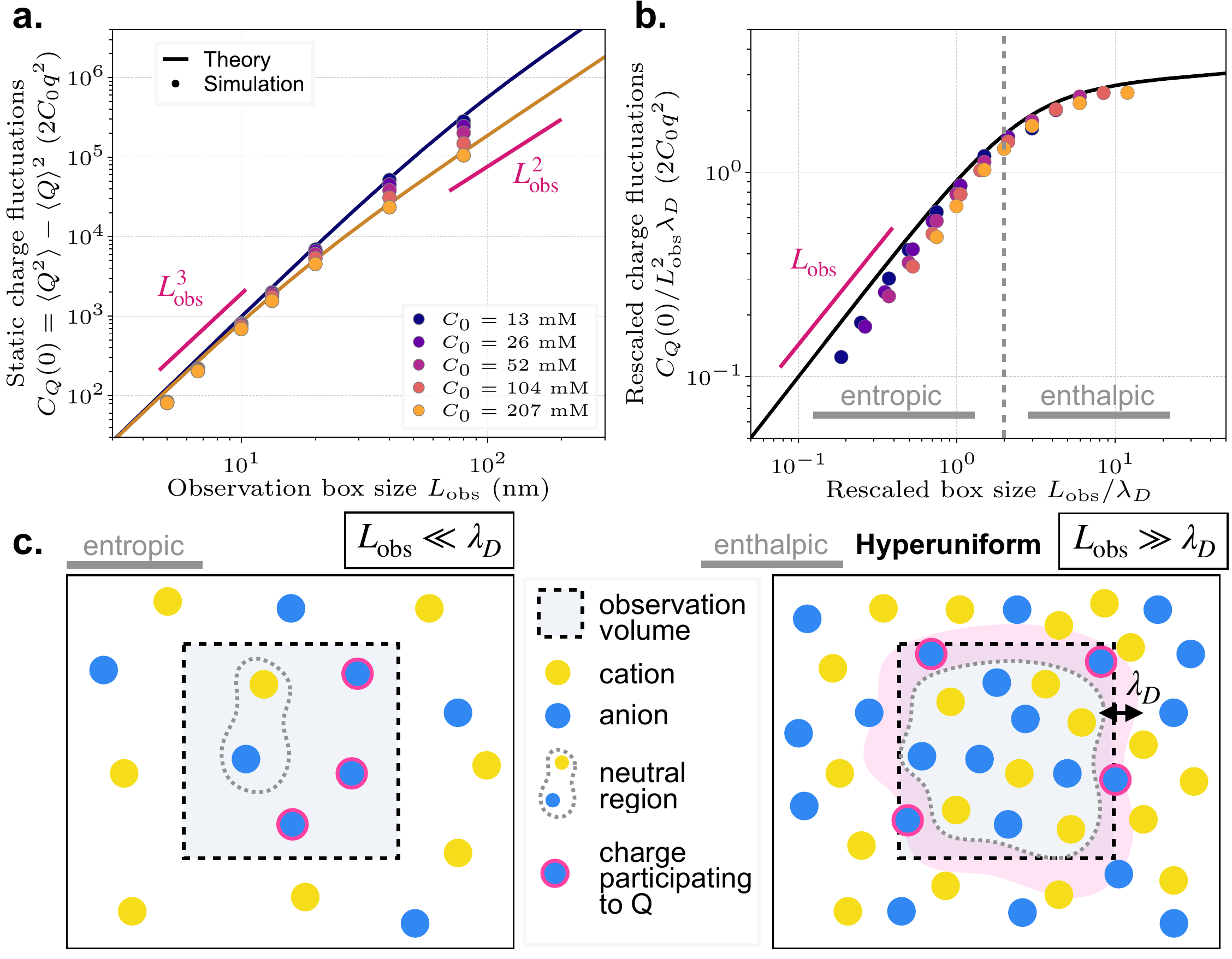}
  \caption{Static charge fluctuations are hyperuniform. (a) Static charge fluctuations with observation box size $\Lbox$ for increasing salt concentrations, going from purple to yellow. Dots: results from BD simulations; lines: Eq.~\eqref{eq:Q20}. Error bars are one standard deviation about the mean and are smaller than dot sizes. (b) Rescaled (a) plot showing data collapse, highlighting ``entropic'' and ``enthalpic'' regimes. (c) Sketch of the origin of fluctuations, in 2D for simplicity, in both regimes (see text for details). Fluctuations growing as the area of the observation volume in the enthalpic regime are, by definition, hyperuniform. }
  \label{fig:fig3}
\end{figure}

We first revisit static charge fluctuations in an observation volume, $\CQ(0) = \langle Q^2 \rangle$, to lay the ground for time dependence. 
For sufficiently small $\Lbox$, BD results indicate that charge fluctuations scale with the volume, $\CQ(0) \sim \Lbox^3$, hence like the average particle number in that region (Fig.~\ref{fig:fig3}-a, circles). In contrast, for large $\Lbox$, and especially at high concentrations, fluctuations scale only with the area of the probe volume, $\CQ(0) \sim \Lbox^2$. This peculiar behaviour was predicted theoretically by \citet{martin1980charge} then verified with Monte Carlo simulations~\cite{lebowitz1983charge, jancovici2003charge, kim2005screening, kim2008charge}. The property that fluctuations over an observation volume scale with the area is nowadays termed \textit{hyperuniformity}, and is a generic feature of particle systems with long-range $1/r$ interactions, where $r$ is the interparticle distance~\cite{torquato2018hyperuniform, ghosh2017fluctuations, leble2021two}.

We may rationalize this behaviour with sDFT: From Eq.~\eqref{eq:deltaNS}, static charge fluctuations integrate the static structure factor as 
\begin{equation}
    \CQ(0) = q^2 \langle N \rangle \int \frac{\dd \bm{k}}{(2\pi)^3} f_\mathcal{V}(\bm{k}) S_{\rho\rho}^{\rm static}(k) = q^2 \langle N \rangle \int \frac{\dd \bm{k}}{(2\pi)^3} f_\mathcal{V}(\bm{k}) \frac{k^2}{k^2 + \kappa_D^2}
    \, ,
    \label{eq:Q20}
\end{equation}
where we recall that $\langle N \rangle = 2C_0 \Lbox^3$ is the average particle number in the box, $\kappa_D = 1/\lambda_D$ and $f_\mathcal{V}(\bm{k})$ is a geometric factor, see Eq.~\eqref{eq:fv}.  $\CQ(0)$ depends only on the ratio $\Lbox/\lambda_D$. Formally, if $\Lbox \ll \lambda_D$ the dominant part of the spectrum will be for values $k \gg \kappa_D$ and expanding Eq.~\eqref{eq:Q20} one obtains $\CQ(0) \sim q^2 \langle N \rangle = 2C_0 q^2   \Lbox^3 $ (see Appendix C). In contrast, when $\Lbox \gg \lambda_D$, further expansions yield $\CQ(0) \sim 2 c_Q C_0  q^2  \Lbox^2 \lambda_D$ where $c_Q = (16\pi)^{1/3} \simeq 3.7$ ($c_Q$ usually depends on box geometry, see~\citet{kim2008charge} and also Appendix D). Fig.~\ref{fig:fig3}-b shows that rescaling $\CQ(0)$ by $\Lobs^2 \lambda_D$ and $\Lobs$ by $\lambda_D$ indeed collapses all results on a master curve. The present framework thus shows that hyperuniformity is not just a property of the system itself but also a property of the observation scale ($\Lobs$) relative to the scale of the interactions ($\lambda_D$). 

This behaviour can further be interpreted with an energetic approach, brought forth by several authors~\cite{van1979thermodynamics, kim2005screening, lebowitz1983charge}. When $\Lobs \ll \lambda_D$, at the observation scale $\Lobs$, electrostatics do not govern ionic structure. Hence, one can simply place particles in the observation volume without concern for their respective charge, among the diversity of particle arrangements (see Fig.~\ref{fig:fig3}-c). Fluctuations are dominated by entropy, and as in any such statistical physics framework, fluctuations scale like the average number of particles in the observation volume. In contrast, when $\Lobs \gg \lambda_D$, an inner, neutral region of the observation volume exists where charges are balanced. The remaining degree of freedom is at the interface, within a thin shell around the neutral region of thickness $\lambda_D$, and fluctuations are dominated by the energetic cost to charge the interface. Hence fluctuations scale as $\Lobs^2 \lambda_D$ and can be viewed as dominated by enthalpy. This entropic/enthalpic interpretation is similar to solute particle fluctuations in a liquid, where the free energy cost to create a spherical cavity scales with the volume for small radii and the area for large radii, with a cross-over around 1~nm in water~\cite{chandler2005interfaces}. Since this free energy cost has proved useful to characterize the hydrophilic/phobic behaviour of interfaces from the local water density fluctuations~\cite{patel2010fluctuations, rotenberg2011molecular, rego_understanding_2022}, it might be relevant to explore this analogy in the case of charged systems further.

\subsection{Charge fluctuations with time: timescales and hyperuniformity}

\begin{figure}[h]
\centering
  \includegraphics[width = 0.8\textwidth]{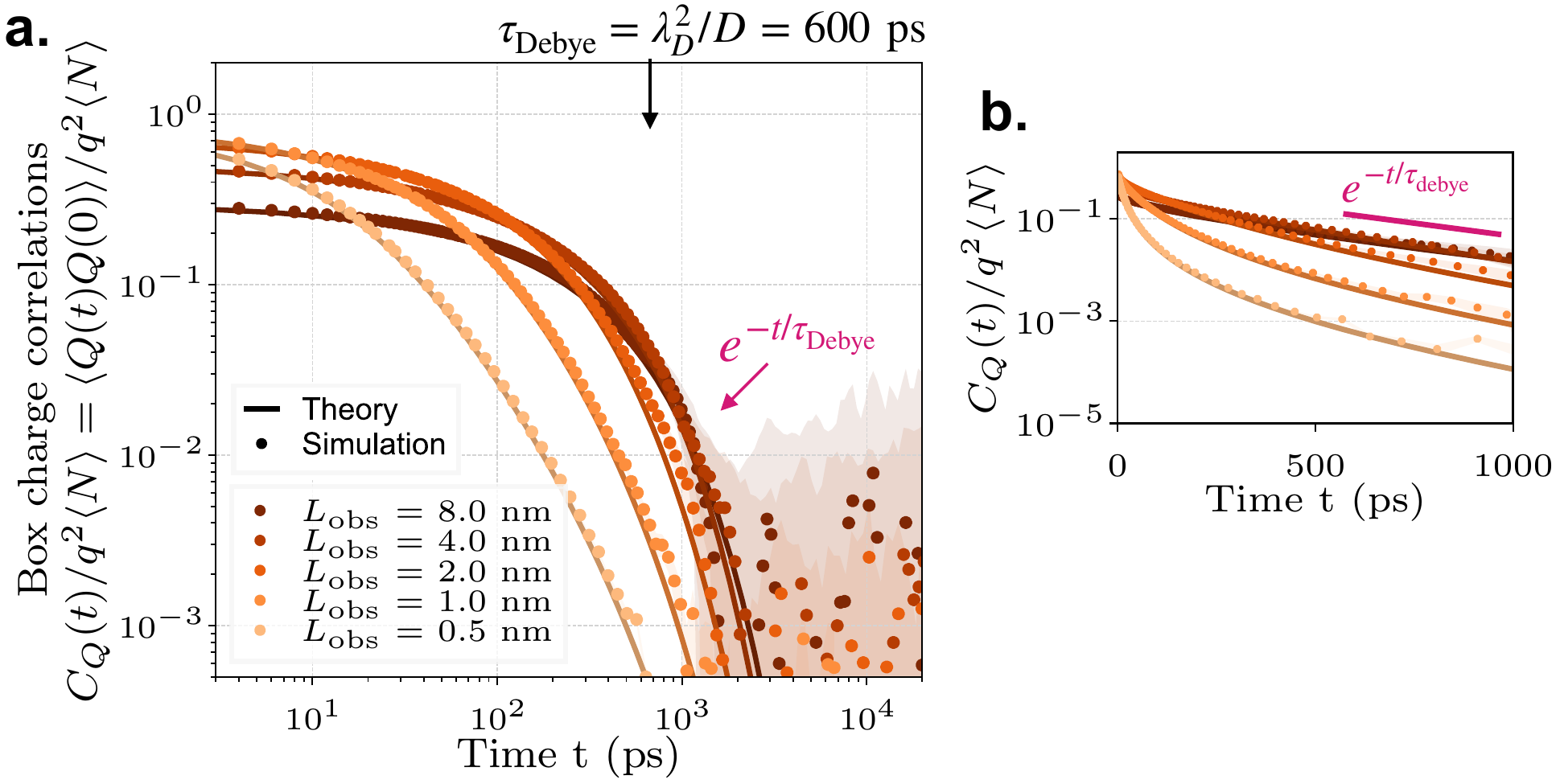}
  \caption{\textbf{Charge fluctuations decay exponentially for large observation volumes.} (a) Charge fluctuations rescaled by $\langle N\rangle$, with time, for increasing $\Lbox$ from yellow to dark red. Dots: results from BD simulations with shaded areas indicating one standard deviation around the mean; lines: Eq.~\eqref{eq:dQt}. (b) Same as (a) in lin-log scale to highlight the exponential decay for large $\Lbox$. Here, $C_0 = 104~$mM and $\lambda_D = 0.95~\mathrm{nm}$. }
  \label{fig:fig4}
\end{figure}

We now turn to the relaxation of charge correlations by considering $\CQ(t) = \langle Q(t) Q(0) \rangle$. Fig.~\ref{fig:fig4}-a displays the BD results for a fixed salt concentration and various observation volumes, rescaled by $q^2\langle N \rangle$. At early times, correlations collapse for small $\Lbox$ (light orange) but not for large ones (dark red), a natural consequence of the above-discussed static ($t=0$) hyperuniformity. Surprisingly, at long times, we observe the opposite behaviour: correlations collapse for large $\Lbox$ but not for small ones. Furthermore, the decay of charge correlations for large $\Lbox$ is not algebraic, in contrast with number fluctuations, but exponential (see Fig.~\ref{fig:fig4}-b). 

\begin{figure}[h!]
\centering
  \includegraphics[width = \textwidth]{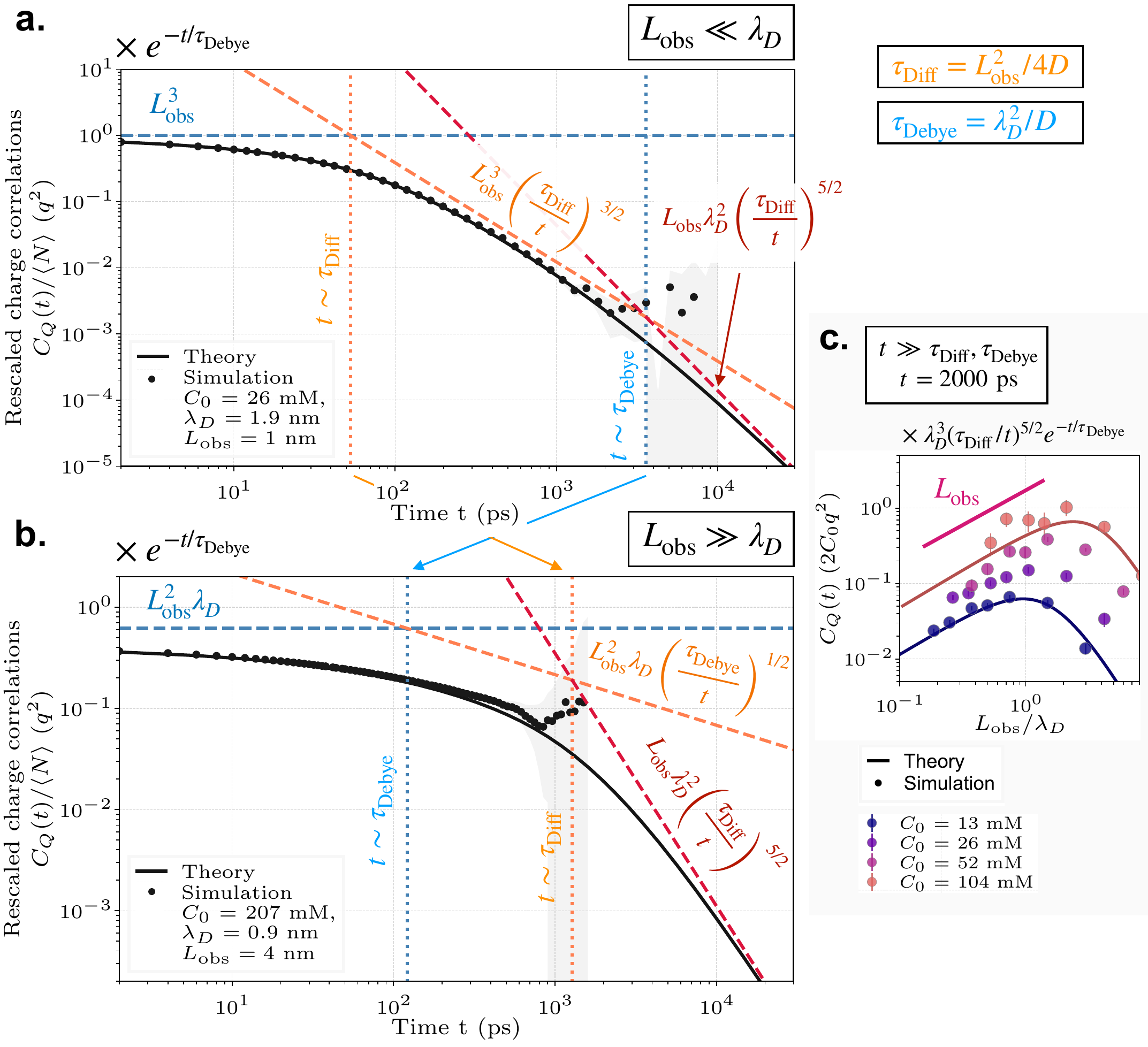}
  \caption{\textbf{A variety of timescales emerges in charge correlation relaxation.} (a) Charge correlations rescaled by $e^{t/\tdeb}$ for $\Lobs \ll \lambda_D$. Vertical dotted orange and blue lines indicate scaling law intersections, as $t = \tdiff/\pi$ and $t = 3\tdeb/2$, respectively. (b) Similar as (a) but for $\Lobs \gg \lambda_D$. The vertical dotted orange and blue lines indicate $t = \sqrt{3}\tdiff/(2^{7/6} \pi^{5/12})$ and $t = 4\tdeb/\pi^2$, respectively. (c) Rescaled charge correlations at long time, with a scaling law as $\CQ \sim \Lobs$.  In all panels: dots: results from BD simulations with shaded areas (or error bars) indicating one standard deviation around the mean; lines: Eq.~\eqref{eq:dQt}. Slight discrepancy arount $t = 1000~\mathrm{ps}$ in (b) between simulations and theory can be attributed to steric effects. 
  }
  \label{fig:fig5}
\end{figure}

We can get direct insight on the exponential decay using sDFT. Integrating Eq.~\eqref{eq:deltaNS} over $\omega$ yields 
\begin{equation}
\begin{split}
    \CQ(t) &=  q^2 \langle N \rangle  \int \frac{d\bm{k}}{(2\pi)^3} S_{\rho \rho}^{\rm static}(k) e^{-D k^2 t / S_{\rho \rho}^{\rm static}(k)}  f_\mathcal{V}(\bm{k}). \\
    &=  q^2 \langle N \rangle \, e^{-t/\tdeb}  \int \frac{d\bm{k}}{(2\pi)^3} \frac{k^2}{k^2 + \kappa_D^2}e^{-D k^2 t}  f_\mathcal{V}(\bm{k})
    \label{eq:dQt}
\end{split}
\end{equation}
where the Debye time $\tdeb = \lambda_D^2/D$ corresponds to the time to diffuse across the Debye length scale. Eq.~\eqref{eq:dQt} reproduces remarkably well the BD results (see Fig.~\ref{fig:fig4}, lines). For sufficiently large $\Lbox$, the correlations decay exponentially with characteristic timescale $\tdeb$. Indeed, the relaxation of charge fluctuations is primarily driven by electrostatics: the transient local breakdown of electroneutrality induces an internal electric field driving the ions (with a mobility $q D/k_BT$) to restore electroneutrality. There are clearly other timescales involved in the relaxation, especially for small $\Lbox$. As mentioned in the introduction, the interplay between $\tdeb$ and $\tdiff$ can produce a variety of  timescales that could all explain part of the behaviour~\cite{bazant2004diffuse, palaia2023charging, minh2022frequency}. To understand the relaxation behaviour more systematically, we explore in Fig.~\ref{fig:fig5} the relaxation of $\CQ(t) e^{t/\tdeb}$. Since BD results are well captured by sDFT over a broad range of parameters, we use analytical expansions of Eq.~\eqref{eq:dQt} to quantify the dependence of the results on $\lambda_D$ and $\Lobs$. 

Fig.~\ref{fig:fig5}-a first reports the case of small observation volumes compared to the Debye length $\Lobs \ll \lambda_D$. Beyond the initial static regime where $\CQ \sim \Lobs^3$, when $t \gtrsim \tdiff$, we find, expanding Eq.~\eqref{eq:dQt}, that the correlations decay as $\CQ \sim \Lobs^3 (\tdiff/t)^{3/2}$ (see Appendix C). This decay exactly follows that of the particle number decay in Fig.~\ref{fig:fig2}-b. At this observation length scale \textit{and} timescale, electrostatics do not play any role, and the only relevant timescale appears to be $\tdiff$. Eventually, at longer times, $t\gtrsim \tdeb$, correlations decay faster as $\CQ \sim \Lobs \lambda_D^2 e^{-t/\tdeb} (\tdiff/t)^{5/2}$, and the Debye timescale $\tdeb$ appears to govern charge fluctuation relaxation. As explained above, this time scale emerges due to restoring electrostatic forces that damp fluctuations arising from diffusion.

How do these effects survive when the length scales, $\Lobs \gg \lambda_D$, and hence the timescales $\tdeb \ll \tdiff$ are reversed? In Fig.~\ref{fig:fig5}-b, we show BD results with parameters $\Lobs \simeq 2 \lambda_D$, which is already hard to achieve with reasonable simulation times. 
Beyond the static hyperuniform regime where $\CQ \sim \Lobs^2 \lambda_D$, for $t\gtrsim \tdeb$ we find $\CQ \sim \Lobs^2 \lambda_D e^{-t/\tdeb} (\tdeb/t)^{1/2}$. The decay of the correlations is apparently entirely due to electrostatic effects, with $\tdeb$ the relevant timescale, and is faster than exponential. Finally, for $t \gtrsim \tdiff$, when particles have had time to diffuse across the observation volume, $\tdiff$ appears in the dynamics, as $\CQ \sim \Lobs \lambda_D^2 e^{-t/\tdeb} (\tdiff/t)^{5/2}$.

Curiously, at long times, correlations decay as $\CQ \sim \Lobs \lambda_D^2 e^{-t/\tdeb} (\tdiff/t)^{5/2}$ in both the $\Lobs \ll \lambda_D$ and $\Lobs \gg \lambda_D$ regimes. At such long timescales, particles have diffused over distances long enough that $\lambda_D$ and $\Lobs$ appear comparably small. Remarkably, the amplitude of the fluctuations now scales with the \textit{perimeter} $\Lobs$ of the observation domain, which we verify numerically in Fig~\ref{fig:fig5}-c. 
Note, that the collapse of the data onto the scaling law is not perfect, since we are limited in time with simulations and the time investigated is not always much bigger than $\tdeb, \tdiff$ for all parameters ($C_0$,$\Lobs$). 
This extreme long-time scaling appears to be a case of hyperuniformity, where the dimensional degree of hyperuniformity is increased because fluctuations have relaxed. It is tempting to interpret this result in the following way: at long times, only boundary crossings in volume elements surrounding cube edges with area $\lambda_D^2$ matter. This open interpretation could be formally addressed, for example, by investigating the spatial relaxation of fluctuations. Finally, with this curious scaling, we might expect that for quasi-2D electrostatics, such as in extremely confined systems~\cite{mouterde2019molecular,robin2023long}, fluctuations should have the same amplitude at long times. This resonates, more generally, with the peculiar behaviour of fluctuations confined to 2D, from thermal Casimir forces to memory effects~\cite{dean2016nonequilibrium,mahdisoltani2021long,robin2023long}.  

Notably, in deriving such scaling laws, there are various ways one can non-dimensionalize time; hence, the relevant timescale is ambiguous. 
For example, at long times, the scaling law for charge correlations can be written in several ways
\begin{equation}
    \CQ(t)  \stackrel[t \gg \tdiff,\tdeb]{}{\sim}  \Lobs \lambda_D^2 \left(\frac{\tdiff}{t}\right)^{5/2} e^{-\frac{t}{\tdeb}} = \Lobs^2 \lambda_D  \left(\frac{\tdiff^{4/5}\tdeb^{1/5}}{t}\right)^{5/2} e^{-\frac{t}{\tdeb}} = ... 
    \label{eq:ambiguity}
\end{equation}
 To interpret these scalings, we generally assume a rule of aesthetic simplicity; at long times, this corresponds to $\CQ \sim \Lobs \lambda_D^2 (\tdiff/t)^{5/2} e^{-t/\tdeb}$. However, this ambiguity highlights the diversity of timescales at play. Together with the variety of scaling laws uncovered in Fig.~\ref{fig:fig5}, this naturally raises the question of understanding which timescale dominates the relaxation of charge fluctuations.

\subsection{Universal timescale to characterize relaxation of fluctuations}
 
We, therefore, define a global, unambiguous, relaxation timescale for charge fluctuations as 
\begin{equation}
    T_Q = \int_0^{\infty} \frac{\CQ(t)}{\CQ(0)}\dd t = \frac{L_Q^2}{D},  \,\,\,\,
    \text{where} \,\,\,\,L_Q^2 =  \frac{\displaystyle \int \frac{d\bm{k}}{(2\pi)^3} \frac{S^{\rm static}_{\rho\rho}(k)^2}{k^2}  f_\mathcal{V}(\bm{k})}{\displaystyle \int \frac{d\bm{k}}{(2\pi)^3} S^{\rm static}_{\rho\rho}(k)  f_\mathcal{V}(\bm{k})}
    \label{eq:T}
\end{equation}
where the last equality comes from Eqs.~\eqref{eq:Q20} and ~\eqref{eq:dQt}.  When calculating $T_N$, using $\textit{i.e.}$ Eq.~\eqref{eq:T} but with the structure factor $S_{cc}^{\rm static}(k) = 1$ for the particle number fluctuations, we find $T_N \simeq \tdiff$, which means that Eq.~\eqref{eq:T} is indeed suited, \textit{a priori}, to uncover a relevant relaxation timescale of the system. Interestingly, introducing $T_Q$ means introducing a characteristic length scale $L_Q$, quantifying how far a particle should diffuse before correlations decay.

\begin{figure}[h]
\centering
  \includegraphics[width = \textwidth]{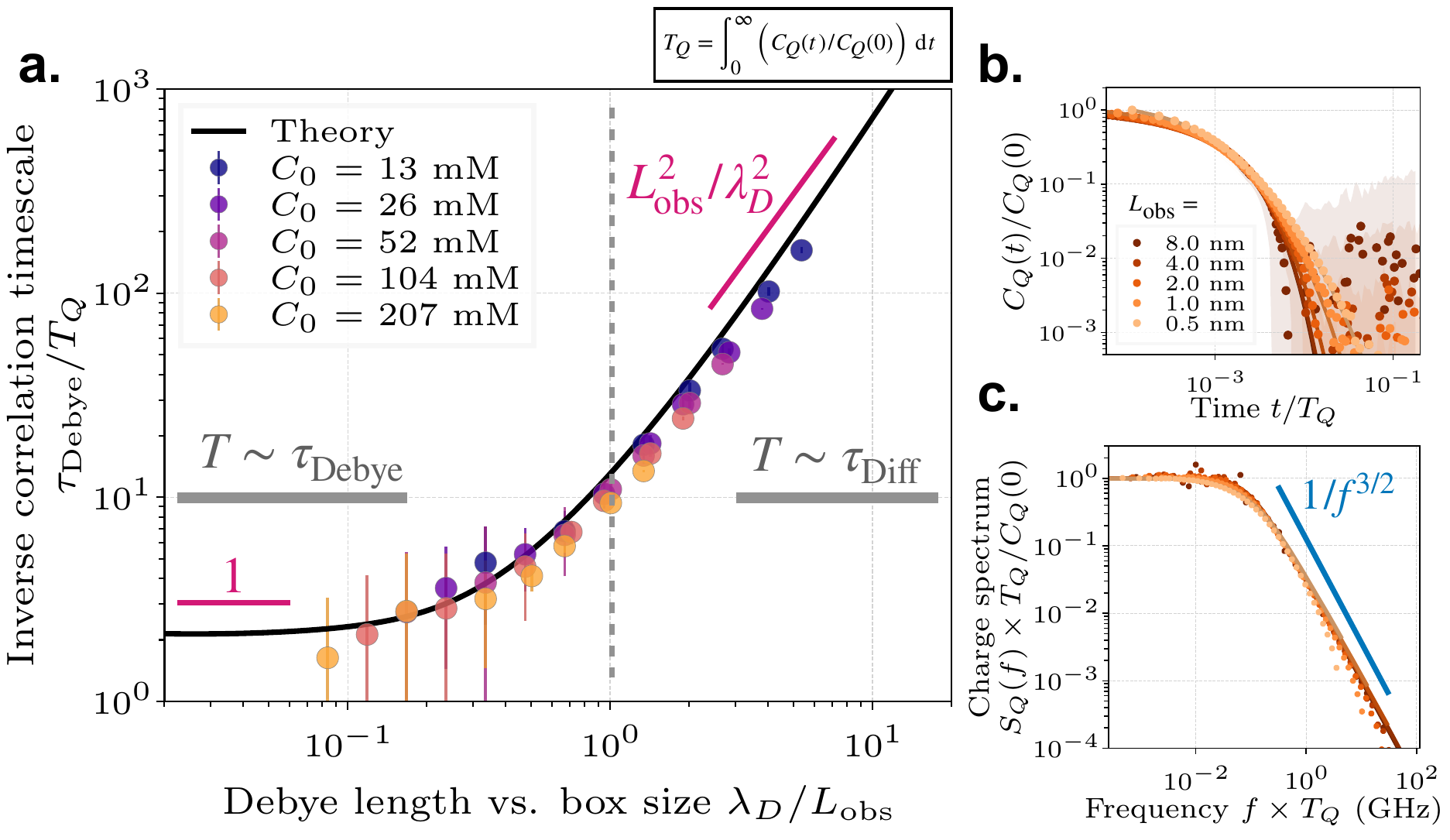}
  \caption{A global timescale $T_Q$, defined by Eq.~\eqref{eq:T}, accounts for relaxation of charge fluctuations. (a) $\tdeb/T_Q$ as a function of $\lambda_D/\Lobs$ for increasing salt concentrations, from purple to yellow. The scaling behaviour of $T_Q$ is highlighted in both limit regimes. Line: Eq.~\eqref{eq:T}. (b) Charge correlations, from Fig.~\ref{fig:fig4}, rescaled by static charge correlations, with time rescaled by $T_Q$. Increasing observation box sizes, from yellow to dark red, for $C_0 = 104$~mM. (c) Power spectrum of (b), as a function of frequency. In all plots, dots: BD data; error bars are one standard deviation around the mean.}
  \label{fig:fig6}
\end{figure}

We plot $T_Q$ as the BD correlations integrated in time in Fig.~\ref{fig:fig6}-a (dots), together with Eq.~\eqref{eq:T} (line).
All BD results for $T_Q$ collapse on a single master curve when presented against one characteristic timescale, here $\tdeb/T_Q$, as a function of the separation of length scales $\lambda_D/\Lobs$. Unsurprisingly, for $\lambda_D \ll \Lobs$, we find $T_Q \sim \tdeb$, while for $\lambda_D \gg \Lobs$, $T_Q \sim \tdiff$, showing that the relevant timescale for the correlations is always the smallest one (see also Appendix C). However, there is a broad intermediate region where $T_Q$ spans a combination of both timescales -- resonating with other studies which also find numerous timescales to characterize the charge of electrodes~\cite{bazant2004diffuse,palaia2023charging,minh2022frequency}.

The global timescale $T_Q$ accounts remarkably for the relaxation of charge correlations. Fig.~\ref{fig:fig6}-b shows that rescaling the time by $T_Q$ collapses the simulation results for normalized charge fluctuations, except for times approaching $T_Q$, where, as we have seen in Eq.~\eqref{eq:ambiguity}, the correlation function 
can only be described with scaling laws involving multiple timescales. The relevance of $T_Q$ is also apparent in the power spectrum of charge fluctuations $S_Q(f)$ rescaled by $T_Q$ (Fig.~\ref{fig:fig6}-c). We find that fluctuations plateau at low frequencies, as $S_Q(f=0) = C_Q(0)/T_Q$, containing the information of static hyperuniformity and relaxation time. The plateau thus corresponds the equilibration of particles inside and outside the observation box at long enough times $t\gtrsim T_Q$. At large frequencies, fluctuations decay as $1/f^{3/2}$, similarly to number fluctuations. This decay occurs for frequencies typically larger than $1/T_Q$, showing that $T_Q$ determines how long correlations persist in the observation box, or determines the length scale, $L_Q = \sqrt{D T_Q}$ to diffuse accross before correlations are lost. Again, the $1/f^{3/2}$ is a signature of fractional noise and shows that this universal behaviour can be seen regardless of the details of pairwise interactions.

The collapse of timescales on a single master curve spanning all intermediate combinations of $\tdeb$ and $\tdiff$ is strikingly similar to the result of a recent study, including some of the authors, of the response to an oscillating electric field of an electrolyte \emph{confined} between two plates separated by a distance $L$~\cite{minh2022frequency}. There, the critical timescale defining a conducting or insulating behaviour, typically the time to ``charge'' the plates by transporting the ions, is either close to $\tdeb$ or $\tau_{\rm Diff} = L^2/D$ according to the separation of length scales $\lambda_D/L$, and spans all intermediate regimes. Remarkably, here in an equilibrium and \textit{bulk} context, this behaviour remains, showing the universality of such response in electrolytes. 

\section{Conclusion and Discussion}
\label{sec:4}

In this work, we have investigated ionic fluctuations in finite observation volumes, in the dilute regime and at equilibrium. With Brownian dynamics simulations and analytical calculations, we have probed the relaxation of correlations in the particle number $N$ and charge $Q$ in the observation volume. For charges, correlations decay with a timescale depending on the separation of length scales between the size of the observation volume $\Lobs$ and the Debye screening length $\lambda_D$: ranging from the time to diffuse across the box $\tdiff = L^2/4D$ to the time to diffuse across the Debye length $\tdeb = \lambda_D^2/D$, and spanning combinations in between. The decay of charge correlations at long times is exponential. In contrast, for particle number, correlations decay algebraically with a single timescale $\tdiff$, independently of electrostatics. We find that charge correlations are hyperuniform when the size of the overvation volume is much larger than the Debye length ($\Lobs \gg \lambda_D$). 
Hyperuniformity persists in time and is even exacerbated at long times, including for small boxes. Finally, both $N$ and $Q$ feature a $1/f^{3/2}$ decay in their power spectrum, a signature of fractional noise, showing the universality of such traces when observing particles diffusing finite volumes. 

\paragraph*{Beyond Debye-H\"uckel: generality of the approach}

Stochastic density functional theory remarkably reproduced simulation results and is applicable to a few more complex systems. In fact, the present analytic theory depends only on the static structure factor of the quantity of interest $X$, $S^{\rm static}_{XX}(k)$. As long as the dynamic structure factor in Eq.~\eqref{eq:Sgeneral} well describes the dynamics of $X$, which is typical for Markovian and Gaussian systems near equilibrium, we can use Eq.~\eqref{eq:dQt} to compute the correlations of $X$ and Eq.~\eqref{eq:T} to characterize their relaxation time. This is especially interesting since the static structure factor $S^{\rm static}_{XX}(k)$ is sometimes hard to calculate analytically but is fairly accessible experimentally and numerically~\cite{FDspectra, thorneywork2018structure}, allowing one to estimate dynamical quantities from static properties. For example, one could explore, in this way, the effect of steric repulsion, which will be the purpose of a further study~\cite{AliceExclusion}. However, significant extensions of sDFT would be required, \textit{e.g.} to model ions with different self-diffusion coefficients, with concentration-dependent diffusion~\cite{dufreche2002ionic}, and with hydrodynamic interactions~\cite{ladiges1,ladiges2}. 

\paragraph*{Extracting kinetic properties from fluctuations in finite volumes.}

Beyond relaxation dynamics, other ionic-specific kinetic properties may be extracted from dynamical fluctuations in finite volumes. For example, conductivity~\cite{zorkot2016power} and the dielectric permittivity and susceptibility~\cite{FDspectra}  also derive from integrals of structure factors, and can be addressed within the same theoretical framework as proposed here.
Since even a quantity as simple as the number density has non-trivial fluctuations, with fractional noise signature, it is clear that more complex variables might exhibit rich behaviour in finite volumes. This resonates with coarse-graining issues, for example, with Lattice methods, where fluctuations do not diminish with coarse-graining size in non-ideal systems, such as with steric repulsion~\cite{parsa2020large,parsa2017lattice}.

\paragraph*{Hyperuniformity in time.}

Here we have highlighted that the hyperuniform behaviour persists in time, reaching peculiar scalings, especially at long times. Yet, electrolytes are but a special case of particles with long-range interactions (decaying as $1/r$ where $r$ is the distance between particles), which include also one-component plasmas, active particles, and many others~\cite{torquato2018hyperuniform, leble2021two, ghosh2017fluctuations}. A recent investigation showed remarkable results where long-range correlations were observed both in driven electrolytes and active particle systems~\cite{mahdisoltani2022nonequilibrium, mahdisoltani2021long}, for the same underlying mathematical reason. This raises the question of whether the time-dependent behaviour uncovered in the present work extends to this broad class of systems and whether other universal signatures may be unraveled.

\paragraph*{Fractional noise and noise in confined systems.} The omnipresence of the spectrum scaling as $1/f^{3/2}$ when observing fluctuations in finite volumes suggests that such fractional noise could be seen in various contexts. Especially in nanopores, one might wonder if fractional noise is linked with the pink noise $\sim 1/f$ scalings measured on current correlations~\cite{secchi2016scaling,bezrukov2000examining,powell2009nonequilibrium,siwy2002origin,knowles2019noise,dekker2007solid,smeets2008noise,knowles2021current}. Beyond apparent discrepancies (in nanopores such correlations are measured out-of-equilibrium), both contexts involve tracking fluctuations in finite sub-volumes of a larger domain~\cite{kavokine2019ionic,gravelle2019adsorption,marbach2021intrinsic,bezrukov2000particle,zorkot2018current,zorkot2016power,zorkot2016current}. The $\sim 1/f$ pink noise arises in more varied electrochemical contexts than nanopores, especially near interfaces, for example in redox monolayers~\cite{grall2022electrochemical}.
More generally, the advent of microscopy techniques resolving electrochemical fluctuations at the single particle level~\cite{zevenbergen_electrochemical_2009, mathwig_electrical_2012,sun_electrochemistry_2008,grall_attoampere_2021,knowles2021current} means that there are increasingly more opportunities to compare experiments and theory at the microscopic level, and more contexts to understand the kinetic response of fluctuations in confined or finite volumes.

\section*{Author Contributions}

\textbf{Th\^e Hoang Ngoc Minh:} Conceptualization (equal); Formal Analysis (equal); Investigation (equal); Validation (equal); Writing/Review \& Editing (equal);
\textbf{Benjamin Rotenberg:} Conceptualization (equal); Formal Analysis (supporting);  Funding Acquisition (lead); Investigation (supporting); Supervision (equal); Validation (equal);  Writing/Review \& Editing (equal).
\textbf{Sophie Marbach:} Conceptualization (lead); Formal Analysis (equal); Funding Acquisition (supporting); Investigation (equal); Supervision (equal); Validation (equal); Writing/Review \& Editing (lead).

\section*{Conflicts of interest}
There are no conflicts to declare.

\section*{Acknowledgements}
We wish to acknowledge fruitful discussions with David S. Dean, Pierre Illien, Thomas Lebl\'{e}, Brennan Sprinkle and Alice Thorneywork.  S.M. received funding from the European Union’s Horizon 2020 research and innovation programme under the Marie Skłodowska-Curie grant agreement 839225, MolecularControl. This project received funding from the European Research Council under the European Union’s Horizon 2020 research and innovation program (project SENSES, grant Agreement No. 863473).

\appendix

\renewcommand\thesection{Appendix~\Alph{section}}

\renewcommand\theequation{\Alph{section}.\arabic{equation}}    
\setcounter{equation}{0}    

\section*{Appendix}

\section{Simulation details}

\paragraph*{Brownian dynamics}

Brownian dynamics simulations were performed using the molecular dynamics simulation code LAMMPS \cite{thompson_LAMMPS_2022} for a typical binary symmetric electrolyte close to aqueous KCl solution at ambient temperature. Periodic boundary conditions in all directions are used to model a bulk electrolyte. The short-range steric repulsion, similar for anions and cations, is modeled by a Weeks Chandler Andersen (WCA) potential described by Eq.~\eqref{eq:WCA_POT}, with parameters $\epsilon_{\mathrm{WCA}} = 0.1$~kcal/mol and $\sigma_{\mathrm{WCA}} = 0.3$~nm. Anions and cations, carrying formal charges $\pm e$, interact through Coulomb interactions Eq.~\eqref{eq:COULOMB_POT} screened by the relative permittivity of water fixed at $\epsilon_\mathrm{w} = 78.5$. Long-range electrostatic interactions are computed with a PPPM  algorithm~\cite{pollock_1996_P3M} with a cutoff distance of $r_\mathrm{cut}= 5\sigma_{\rm WCA} = 1.5$~nm and a relative force error of $10^{-5}$. The diffusion coefficients, set to $D_+ = D_- = D =  1.5~10^{-9}$ m$^2$/s, parametrize the random force and the mobility through fluctuation-dissipation $ D = \mu k_B T$, with $T=300$K. Equations of motion are solved using an adapted overdamped BAOAB integrator~\cite{Leimkuhler_BAOAB_2013} with a timestep of $\delta t = 20$ fs. 

\begin{equation}
\begin{split}
    V^\mathrm{WCA}(r) &=
\begin{cases}   V^\mathrm{LJ}(r) \, - \, V^\mathrm{LJ}(2^{1/6} \sigma_\mathrm{WCA}) \ &, \ r \leq 2^{1/6} \sigma_\mathrm{WCA}, \\
    0 \ &, \ r > 2^{1/6} \sigma_\mathrm{WCA},
    \end{cases}
     \\
   & \text{where} \,\,    V^\mathrm{LJ}(r) = 4 \epsilon_\mathrm{WCA} \left[ \left( \frac{\sigma_\mathrm{WCA}}{r} \right)^{12} - \left( \frac{\sigma_\mathrm{WCA}}{r} \right)^{6}\right]
    \end{split}
\label{eq:WCA_POT}
\end{equation}

We run simulations keeping the simulation box size fixed $\Lsim = 16~\mathrm{nm}$ while changing the number of particles to modulate salt concentration. We recapitulate in Table~\ref{tab:systems} for each simulation system the resulting salt concentration $C_0$, Debye length  $\lambda_D$ and the relative volume packing fraction. Overall, we work at dilute salt concentrations, meaning that packing fractions are low enough that we may neglect steric effects~\cite{thorneywork2018structure}.

\begin{table}[ht!]
    \centering
  \begin{tabular}{cccc}
    $N_0$ & $C_0$ (mM) & $ \lambda_D$ (nm) & packing fraction \\
    \hline
    32 & 13 & 2.7 & 0.17 \%\\
    64 & 26 & 1.9 & 0.35 \%\\
    128 & 52 & 1.3 & 0.71 \%\\
    256 & 104 & 0.95 & 1.4 \% \\
    512 & 207 & 0.67 & 2.8 \% \\
  \end{tabular}
    \caption{Main physical parameters for all simulated systems: number of ion pairs, salt concentration, Debye screening length and packing fraction.}
    \label{tab:systems}
\end{table}

\paragraph*{Data analysis}

For each simulation, we store particle positions every $\Delta t = 100 \delta t$ and run simulations for a total of $N_T = 5 \times 10^7$ time steps (except for $C_0 = 207$~mM for which $N_T = 2.5 \times 10^7$ time steps). We then track particle locations in various boxes through a custom-made Python routine. The simulation box is then divided into $N_{\rm bin}^3$ cubic observation volumes with size $\Lobs = \Lsim/N_{\rm bin}$ where $N_{\rm bin} = [2,4,8,12,16,24,32]$ are used. We then average correlations over every single recorded time step as $\CQ(t = n\Delta t) = \frac{1}{N_T - n}\sum_{p=0}^{N_T - n} Q(t_0 = p\Delta t)Q(t_0+ t =(p+n)\Delta t)$. The uncertainty is estimated as the standard deviation of samples $Q(t_0 = p\Delta t)Q(t_0+ t =(p+n)\Delta t)$ divided by the square root of the number of totally uncorrelated samples, which is $(N_T - n)/n$ here. 

\paragraph*{Power spectrum}

For a given random variable $X$ we define its power spectrum as 
\begin{equation}
S_X(f) = \lim_{\tau\rightarrow \infty} \frac{1}{\tau} \bigg| \int_0^{\tau} e^{i 2\pi f t} X(t) dt\bigg|^2 = \lim_{\tau\rightarrow \infty} \int_0^{\tau} e^{i 2\pi f t} \langle X(t) X(0) \rangle dt
\label{eq:powerSpectrum}
\end{equation}
and is essentially the Fourier transform of the correlation function. $S_X(f)$ is calculated from BD data through fast Fourier transforms of the correlation signal by padding the signal with its time-reversed signal to limit finite acquisition time effects on the Fourier transform, similarly as in~\citet{marbach2021intrinsic}.

\section{Stochastic density functional theory for electrolytes}

The N-body problem Eq.~\eqref{eq:Langevin}, can be rigorously mapped to stochastic density equations using the formalism of sDFT \cite{dean1996langevin}. For our case of binary symmetric electrolyte, the exact equations for the anion and cation densities $c_{\pm}(\bm{r},t)$ read: 
\begin{equation}
\begin{cases}
        \partial_t c_{\pm}(\bm{r},t) + \bm{\nabla} \cdot \bm{J}_\pm (\bm{r},t) = 0\\
        \bm{J}_\pm (\bm{r},t) = -D \left( \bm{\nabla} c_{\pm}(\bm{r},t)  \pm  q c_{\pm}(\bm{r},t) \frac{\bm{\nabla} \phi(\bm{r},t)}{\kT}\right) - \sqrt{2 D \, c_{\pm}(\bm{r},t)} \bm{\eta}_{\pm}(\bm{r},t)
\end{cases}
\end{equation}
where the electrostatic potential $\phi$ satisfies the Poisson equation $\Delta \phi = - q \frac{c_{+}-c_{-}}{\epsilon_w}$ and the 3D noise fields $\bm{\eta}_{\pm}(\bm{r},t)$ are uncorrelated Gaussian noises, with zero averages and $\langle \eta_{\pm,i}(\bm{r},t)\eta_{\pm,j}(\bm{r}',t') \rangle = \delta_{\pm\pm} \delta_{ij} \delta(\bm{r} - \bm{r}') \delta(t-t')$. Analytical solutions of these coupled nonlinear SDEs with multiplicative noises are very limited \cite{te_vrugt_classical_2020}. Here, we briefly summarize the derivation of the dynamic structure factor of the fields, following \citet{mahdisoltani2021long}, where more general cases are treated. 
We switch the formulation to number $C = c_+ + c_-$ and charge $\rho = q (c_+ - c_-)$ densities, giving
\begin{equation}
\begin{cases}
           \partial_t C &= D \nabla^2 C + q  \bm{\nabla} \cdot \left( \rho \frac{\bm{\nabla} \phi(\bm{r},t)}{\kT}\right)  +  \left( \sqrt{4 D C} \bm{\eta}_{c} \right) \\
        \partial_t \rho &= D \nabla^2 \rho + q  \bm{\nabla} \cdot \left(C \frac{\bm{\nabla} \phi(\bm{r},t)}{\kT}\right)  + \bm{\nabla}\cdot  \left( \sqrt{4 D C} \bm{\eta}_{\rho} \right) 
\end{cases}
\end{equation}
where the noise fields $\sqrt{4 D C} \bm{\eta}_{c,\rho}$ are obtained from addition and subtraction of $\sqrt{2 D \, c_{\pm}(\bm{r},t)} \bm{\eta}_{\pm}$, and one can check that the $ \bm{\eta}_{c,\rho}$ are uncorrelated gaussian noise fields, with similar properties as the $\bm{\eta}_{\pm}$. 
We then linearize around the homogenous cation and anion densities (which are equal by electroneutrality), assuming $|c_+- C_0| \sim |c_- - C_0| \ll C_0$ or equivalently $|c = C - 2C_0| \sim |\rho/q| \ll C_0$, and make use of Poisson's equation, which yields Eq.~\eqref{eq:L-sPNP}. The solution is conveniently expressed using spatial and temporal Fourier transforms. This leads to
\begin{equation}
\begin{cases}
     \tilde{c}\left( \bm{k},\omega\right) &= \displaystyle i\frac{\sqrt{4 D C_0}}{-i \omega + D k^2 } \, \bm{k} \cdot\tilde{\bm{\eta}}_c \\
    \tilde{\rho}\left( \bm{k},\omega\right) &=  \displaystyle  i\frac{ \sqrt{4 D C_0}}{-i \omega + D\left(k^2+\frac{1}{\lambda_D^2}\right) }\,  \bm{k} \cdot \tilde{\bm{\eta}}_{\rho}
\end{cases} 
\label{eq-L-sPNP-Fourier}
\end{equation}
where the Fourier transforms of the random fields obey:
\begin{equation}
    \left\langle \tilde{\eta}_{X,i} \left(\bm{k},\omega \right) \tilde{\eta}_{Y,j}^\dagger \left(\bm{k}',\omega'\right) \right\rangle \, = \,
     \delta_{ij} \delta_{XY} \left(2\pi\right)^4\, \delta^3(\bm{k}'+\bm{k}) \delta(\omega'+\omega)
\label{eq:random_fields_fourier}
\end{equation}
for $X = \in\{c,\rho\}$ and $i \in\{x, y, z\}$. Finally, computing the quantities $\left\langle \tilde{X}\left( \bm{k},\omega\right) \tilde{X}^\dagger\left( \bm{k}',\omega'\right) \right\rangle$ and using Eq. \eqref{eq:S_fourier} yields the dynamic structure factors Eq.\eqref{eq:Sgeneral}.

\section{Additional calculations of correlation functions}

\paragraph*{Volume factor in Fourier space}

For our cubic observation volume of side $\Lbox$, 
\begin{equation}
\begin{split}
 f_\mathcal{V}(\bm{k}) &=  \frac{1}{\Lbox^3} \iint \dd \bm{x} \dd \bm{x}' e^{i \bm{k}\cdot (\bm{x}- \bm{x}')} \\
 &=  \Lbox^3 \left( \frac{ \sin\left( k_x \Lbox/2 \right) }{k_x \Lbox/2}\right)^2 \left( \frac{ \sin\left( k_y \Lbox/2 \right) }{k_y \Lbox/2}\right)^2 \left( \frac{ \sin\left( k_z \Lbox/2 \right) }{k_z \Lbox/2}\right)^2.
 \end{split}
 \label{eq:fv2}
\end{equation}

\paragraph*{Particle number correlations}

For particle number correlations, we have, starting from Eq.~\eqref{eq:deltaNS} carrying first the integral over $\omega$
\begin{equation}
\begin{split}
    \CN(t) &= \Lbox^3 \iint \frac{d\bm{k}d\omega}{(2\pi)^4} e^{i\omega t} f_\mathcal{V}(\bm{k}) \frac{4DC_0 k^2}{\omega^2 + (Dk^2)^2} \\
    &= \langle N \rangle  \int \frac{d\bm{k}}{(2\pi)^3} e^{-D k^2 t}  f_\mathcal{V}(\bm{k}).
\end{split}
\end{equation}
Observe that $ e^{-D k^2 t} = e^{-D k_x^2 t} e^{-D k_y^2 t} e^{-D k_z^2 t} $ and similarly from Eq.~\eqref{eq:fv2} it is clear that $f_\mathcal{V}(\bm{k})$ can be decomposed over each component of $\bm{k}$. Hence, we can split the integral into 3 separate integrals, 
\begin{equation}
    \CN(t) = \langle N \rangle  \left[\int \frac{dk \Lbox}{2\pi} e^{-D k^2 t} \left( \frac{ \sin\left( k \Lbox/2 \right) }{k \Lbox/2}\right)^2 \right]^3.
\end{equation}
With the change of variable $K = k\Lbox/2$ and recalling that $\tau_{\rm Diff} = \Lbox^2/4D$. we find
\begin{equation}
    \CN(t) = \langle N \rangle  \left[\int \frac{dK}{\pi} e^{-K^2 t/\tau_{\rm Diff}} \left( \frac{ \sin\left(K\right) }{K}\right)^2 \right]^3.
\end{equation}
Carrying out the integral, we obtain Eq.~\eqref{eq:deltaN}.

\paragraph*{Charge correlations}

For charge correlations, we have, starting from Eq.~\eqref{eq:deltaNS} carrying first the integral over $\omega$
\begin{equation}
\begin{split}
    \CQ(t) &= \Lbox^3 \iint \frac{d\bm{k}d\omega}{(2\pi)^4} e^{i\omega t} f_\mathcal{V}(\bm{k}) \frac{4DC_0 k^2}{\omega^2 + (D(k^2+ \kappa_D^2))^2} \\
    &= \langle N \rangle  \int \frac{d\bm{k}}{(2\pi)^3} \frac{k^2}{k^2 + \kappa_D^2}e^{-D (k^2 + \kappa_D^2) t}  f_\mathcal{V}(\bm{k}).
\end{split}
\end{equation}
Unfortunately, in the charge case, for a cubic geometry, it is not possible to split contributions and obtain an explicit expression.
\paragraph*{Approximate expression of the volume factor}
To make progress, we derive an approximate analytic expression for $f_\mathcal{V}(\bm{k})$. Expressing $\bm{k}$ in the spherical coordinate system and with the change of variable $K = k \Lbox/2$, the charge correlations simplify to
\begin{equation}
    \CQ(t) = \langle N \rangle  \int \frac{K^2dK}{\pi^3} \frac{K^2}{K^2 + K_D^2}e^{- (K^2 + K_D^2) t/\tau_{\rm Diff}}  f_V(K).
    \label{eq:Q2ta1}
\end{equation}
where $K_D = \frac{\Lbox}{2 \lambda_D}$ and
\begin{equation}
    f_V(K) =  \iint \sin \theta d \theta d\phi \left( \frac{ \sin(K  \sin \theta \cos \phi)}{K \sin \theta \cos \phi} \right)^2 \left( \frac{\sin(K \sin \theta \sin \phi)}{K\sin \theta \sin \phi} \right)^2 \left( \frac{ \sin(K \cos \theta)}{ K\cos \theta} \right)^2 
\end{equation}
Notice in the above expressions that $K$ is a non-dimensional wave number. When $K \ll 1$, 
\begin{equation}
        f_V(K) \simeq 8  \int_{\theta = 0}^{\pi/2}\int_{\phi = 0}^{\pi/2}  \sin \theta d \theta d\phi \simeq 4\pi .
\end{equation}
When $K\gg 1$, the terms $\sin (K ...)^2$ oscillate very quickly. Integration on $\phi$ (resp. on $\theta$) actually corresponds to $K$ integrations of a finite quantity. This means that $f(K) \sim  K^2/K^6 = 1/K^4$ when $K\gg1$. Overall we can therefore approximate the volume factor by
\begin{equation}
    f_V(K) = \frac{4 \pi}{1 + \alpha K^4}
    \label{eq:fapprox}
\end{equation}
where $\alpha$ is a numerical prefactor.
Rather than looking at a detailed expansion for $K \gg 1$ to obtain $\alpha$, we take $\alpha$ such that we recover the unscreened limit $\lambda_D \gg \Lbox$ ($K_D \ll 1$) where Eq.~\eqref{eq:Q2ta1} is analytically solvable, which requires $\alpha = (2/\pi^2)^{2/3}$. One can then check through numerical integration that Eq.~\eqref{eq:fapprox} is indeed a good approximation of $f_V(K)$ (not shown here). 
The charge correlation function is thus simply 
\begin{equation}
    \CQ(t) \simeq  q^2 \langle N \rangle  \int \frac{4 K^2dK}{\pi^2} \frac{K^2}{K^2 + K_D^2} \frac{1}{1 + \alpha K^4} e^{- (K^2 + K_D^2) t/\tau_{\rm Diff}}.
    \label{eq:Q2ta}
\end{equation}

\paragraph*{Limit regimes for static charge correlations}
At steady state one can easily integrate Eq.~\eqref{eq:Q2ta} and obtain
\begin{equation}
    \CQ(0) =q^2  \langle N \rangle \left( 1 - \frac{1}{1 + 2(2\pi)^{1/3} \frac{\lambda_D}{\Lbox} + 2(2\pi)^{2/3} \frac{\lambda_D^2}{\Lbox^2} }\right)
\end{equation}
for which one recovers easily that when $\lambda_D \gg \Lbox$, $\CQ(0) = q^2 \langle N \rangle $ and when $\lambda_D \ll \Lbox$, $\CQ(0) = q^2 \langle N \rangle 2 (2\pi)^{1/3}\lambda_D /\Lbox$ given in the main text. 

\paragraph*{Limit regimes for time dependence of charge correlations}
We inspect first the case $\Lbox \ll \lambda_D$, which in Eq.~\eqref{eq:Q2ta} corresponds to $K_D \ll 1$. At very short times, where $t \ll \tdiff$, there is no notable variation, and since $\Lbox\ll \lambda_D$, it is clear that
\begin{equation}
    \CQ(t \ll \tdiff) \simeq \CQ(0) =  q^2 N = q^2 (2 C_0) \Lbox^3. 
\end{equation}
When $t \gtrsim \tdiff$, but still $t \lesssim \tdeb$, we may still assume in Eq.~\eqref{eq:Q2ta} that the dominant $K$ values are for $K \gg K_D$ and since $\CQ(t)$ doesn't depend on $\lambda_D$ yet, we can carry out the integral, and we obtain a similar behaviour as for the particle number
\begin{equation}
\begin{split}
    \frac{\CQ(\tdeb \geq t \geq \tdiff)}{q^2\langle N \rangle} &= \left[ f_N\left(\frac{t}{\tdiff} \right)  \right]^3,  \\
   & \text{where}\,\, f_N\left(\tau = \frac{t}{\tdiff} \right) =   \sqrt{\frac{\tau}{\pi}} \left( e^{-1/\tau} - 1\right) + \mathrm{erf} \left( \sqrt{\frac{1}{\tau}} \right). 
    \label{eq:deltaN2}
    \end{split}
\end{equation}
such that if we expand at early times, we obtain
\begin{equation}
\begin{split}
    \frac{\CQ(\tdeb \geq t \geq \tdiff)}{q^2\langle N \rangle} &\simeq  \left(\frac{t}{\pi \tdiff}\right)^{-3/2}.
    \label{eq:deltaN3}
    \end{split}
\end{equation}
Finally, when $t \gg \tdeb$, then only small values of $K \ll K_D$ will contribute to the integral and we can approximate it as
\begin{equation}
    \CQ(t) \simeq q^2 \langle N \rangle  \int \frac{4 K^2dK}{\pi^2} \frac{K^2}{K_D^2} e^{- (K^2 + K_D^2) t/\tau_{\rm Diff}}
    \, ,
    \label{eq:C21}
\end{equation}
which yields
\begin{equation}
   \CQ(t \gg \tdeb) \simeq q^2 (2 C_0) \Lbox \lambda_D^2 \frac{6}{\pi^{3/2}} \left(\frac{t}{\tdiff}\right)^{-5/2} e^{-t/\tdeb}
   \, . 
       \label{eq:tscale3}
\end{equation}

We now study the time dependence when $L \gg \lambda_D$, so in Eq.~\eqref{eq:Q2ta}, this means $K_D \gg 1$. At very short times, where $t \ll \tau_{\rm Debye} = \lambda_D^2/D$ there is no notable variation in time yet and 
\begin{equation}
    \CQ(t \ll \tdeb) \simeq \CQ(0) = (16\pi)^{1/3} q^2 (2 C_0) \Lbox^2 \lambda_D . 
\end{equation}
When $t \gtrsim \tdeb$ yet $t \lesssim \tdiff$, then we can approximate Eq.~\eqref{eq:Q2ta} assuming $e^{-K^2 t/\tdiff} \simeq 1$ when $K \leq \sqrt{\tdiff/t}$ and $0$ otherwise. This leads (assuming first $K_D \gg K$ in Eq.~\eqref{eq:Q2ta}) to
\begin{equation}
    \CQ(t) \simeq q^2 \langle N \rangle  \int_0^{\sqrt{\tdiff/t}} \frac{4 K^2dK}{\pi^2} \frac{K^2}{K_D^2} \frac{1}{1 + \alpha K^4} 
\end{equation}
and expanding for sufficiently long times compared to $\tdeb$,
\begin{equation}
    \CQ(\tdiff \geq t \geq \tdeb) \simeq (16\pi)^{1/3} q^2 (2 C_0) \Lbox^2 \lambda_D \frac{2}{\pi} \sqrt{\frac{\tdeb}{t}} e^{-t/\tdeb}.
    \label{eq:tscale2}
\end{equation}
Finally, when $t \gg \tdiff$, then only small values of $K$ will contribute to the integral, which we can approximate (similarly to the case $\Lbox \ll \lambda_D$) as in Eq.~\eqref{eq:C21} which ultimately yields Eq.~\eqref{eq:tscale3}.

\paragraph*{Charge correlation timescale}

The total correlation timescale can be defined as
\begin{equation}
    T = \int_0^{\infty} dt \frac{\CQ(t) }{\CQ(0) } = \int_0^{\infty} dt \frac{\langle Q(t) Q(0) \rangle }{\langle Q(0) Q(0) \rangle }.
\end{equation}
It can be easily calculated from Eq.~\eqref{eq:Q2ta} as
\begin{equation}
    T =  \tau_{\rm Diff} \frac{  \int K^2 dK  \frac{1}{K^2} \left(\frac{K^2}{K^2 + K_D^2} \right)^2 \frac{1}{1 + \alpha K^4} }{ \int K^2 dK \frac{K^2}{K^2 + K_D^2} \frac{1}{1 + \alpha K^4}  } =  \tau_{\rm Diff} f_T (\Lobs/2\lambda_D) \, ,
    \label{eq:Texpression}
\end{equation}
where 
$$f_T(x) = \frac{\sqrt{\alpha} \left( \sqrt{2} - \alpha^{1/4} x  + \alpha^{3/4} x^3 \right)}{ \left( \sqrt{2} + 2 \alpha^{1/4} x \right) \left( 1 + \alpha x^4 \right)} \, ,
$$ and we recall $\alpha = (2/\pi^2)^{2/3}$. Note again that since $K$ is non-dimensional, Eq.~\eqref{eq:Texpression} is indeed homogeneous.
When $\Lobs \gg \lambda_D$ then 
\begin{equation}
    T = \tdiff \frac{1}{2 (\Lobs/2\lambda_D)^2} = \frac{1}{2} \tdeb.
\end{equation}
Otherwise when $L \ll \lambda_D$
\begin{equation}
    T = \tdiff (2/\pi^2)^{1/3}.
\end{equation}

\paragraph*{Power spectrum}

From the expression of the power spectrum Eq.~\eqref{eq:powerSpectrum} and Eq.~\eqref{eq:Q2ta} we obtain the power spectrum for the charge in the observation volume as 
\begin{equation}
    S_Q(f = \omega/2\pi) = \tdiff \langle N \rangle \int \frac{K^2 dK}{\pi^3} \frac{K^2}{(\omega \tdiff)^2 + (K^2 + K_D^2)^2 } f_V (K)
    \label{eq:Sq}
\end{equation}
and note that $S_N(f) = \lim_{K_D \rightarrow 0} S_Q(f)$. In general, note that 
\begin{equation}
    S_X(f = 0)  = \frac{T_X}{C_X(0)} \, ,
\end{equation} which means the time scale for relaxation and the hyperuniform behaviour are all contained in the zero frequency limit of the power spectrum. We use Eq.~\eqref{eq:Sq} (with the exact $f_V(K)$) to plot lines in Fig.~\ref{fig:fig2}-c and Fig.~\ref{fig:fig6}-c. 
With the approximate expression for $f_V(K)$, we can first obtain an expression for the particle number spectrum
\begin{equation}
    S_N(f = \omega/2\pi) \simeq \langle N \rangle \frac{\tdiff \pi}{2\sqrt{2} \alpha^{1/4}} \frac{1- (\alpha^{1/2} \omega \tdiff)^{1/2}}{1 - (\alpha^{1/2} \omega \tdiff)^{2}} .
\end{equation}
Clearly for small frequencies, the power spectrum plateaus as $S_N(f \ll 1/\tdiff)  = \frac{T}{C_N(0)} \simeq \frac{\tdiff \pi}{2\sqrt{2} \alpha^{1/4}}$. The power spectrum decays indeed as $1/f^{3/2}$ for large frequencies. 
We do not report the approximate expression of $S_Q(f)$ since it is quite lengthy, but similarly, as $S_N(f)$, it can be expanded for small and large frequencies to yield a plateau and a $1/f^{3/2}$ decay, respectively. 

\section{Expressions for a spherical observation volume}

For a spherical observation volume of radius $\Robs$, 
\begin{equation}
    f_{\mathcal{V}}(\bm{k}) = \frac{12\pi \Robs^3}{(k\Robs)^4}\left( \frac{\sin(k\Robs)}{k\Robs}  - \cos(k\Robs) \right)^2 
\end{equation}
Then writing $K = k \Robs$ and $K_D = \Robs/\lambda_D$ and $\tdiff = D/\Robs^2$ we simply have the charge correlation function
\begin{equation}
    \CQ(t) = q^2 \langle N \rangle  \frac{6}{\pi} \int  \frac{K^2}{K^2 + K_D^2}e^{- (K^2 + K_D^2) t/\tau_{\rm Diff}} \frac{1}{K^4} \left(\frac{\sin(K)}{K} - \cos(K)\right)^2 K^2dK \, .
    \label{eq:Q2taR}
\end{equation}
While the integration of Eq.~\eqref{eq:Q2taR} can be done analytically, it yields a rather lengthy result which we do not report here. 

At $t = 0$, we have
\begin{equation}
    \CQ(0) = 3 q^2 \langle N \rangle \frac{\lambda_D (\lambda_D+\Robs) }{\Robs^2} \left( \cosh \left( \frac{\Robs}{\lambda_D}\right) - \frac{\lambda_D}{\Robs}\sinh \left( \frac{\Robs}{\lambda_D}\right)  \right) e^{-\Robs/\lambda_D}  \, ,
\end{equation}
which is exactly the expression obtained in Eq.~(17) of \citet{kim2008charge}.
When $\Robs \gg \lambda_D$, the static charge correlations are indeed hyperuniform
\begin{equation}
     \CQ(0) = \frac{3}{2} q^2 \langle N \rangle \frac{\lambda_D}{\Robs} = 3 q^2 C_0 \Robs^2 \lambda_D.
\end{equation}
In the opposite limit where $\lambda_D \gg \Robs$ we have
\begin{equation}
     \CQ(0) =q^2\langle N \rangle
\end{equation}
as expected from standard statistical physics frameworks. 

Now when $\Robs \gg \lambda_D$ and when we observe the dynamics at early times, here meaning $t \geq \tdeb = \lambda_D^2/D$ but $t \leq \tdiff = \Robs^2/D$, we have
\begin{equation}
    C_Q(\tdiff \geq t \geq \tdeb) \simeq q^2 \frac{3}{\pi} \langle N \rangle  \frac{\lambda_D}{\Robs}  e^{-t/\tdeb}\sqrt{\frac{\tdeb}{t}} \, ,
\end{equation}
so that the dynamics appear to be governed by the Debye time. Notice that the scaling law for the dynamics is very similar to that in the cubic case, Eq.~\eqref{eq:tscale2}.
Finally, at long times, regardless of the separation of length scales
\begin{equation}
    C_Q(t \gg \tdeb, \tdiff) \simeq q^2 \frac{ 32 C_0  \lambda_D^2 \Robs }{\pi} e^{- t/\tdeb} \left( \frac{\tdiff}{t} \right)^{5} e^{-t/\tdeb} \, ,
\end{equation}
which shows again that timescales are convoluted at long times. The expression we obtain is very similar to Eq.~\eqref{eq:tscale3}, except with a $1/t^5$ instead of a $1/t^{5/2}$ scaling in time. Noteworthy, at long times, the correlations appear to scale with $\Robs$, so with the perimeter of the observation volume, which we also found in the box geometry. 

Note, that also in this spherical geometry, we can calculate the Fourier spectrum -- which we do not report here as it is very similar to the steps above -- and we obtain a plateau and a $1/f^{3/2}$ decay at long times. 

%


\balance



\providecommand*{\mcitethebibliography}{\thebibliography}
\csname @ifundefined\endcsname{endmcitethebibliography}
{\let\endmcitethebibliography\endthebibliography}{}

\end{document}